\documentclass[journal,comsoc]{IEEEtran}

\usepackage{cite}
\usepackage[caption=false,font=footnotesize,labelfont=sf,textfont=sf]{subfig}
\usepackage{balance}
\usepackage{tabularx}
\usepackage{booktabs}
\usepackage{amsmath,amssymb,amsfonts}
\usepackage{mathtools} 
\usepackage{graphicx}
\usepackage{xcolor}
\usepackage{subfig}
\usepackage{tikz}
\usepackage{float}
\usepackage[nolist,nohyperlinks]{acronym}
\usepackage{algorithm}
\usepackage{algpseudocode}
 \usepackage{setspace}
\let\Algorithm\algorithm
\renewcommand\algorithm[1][]{\Algorithm[#1]\setstretch{1.1}}

\DeclarePairedDelimiter\abs{\lvert}{\rvert}
\DeclarePairedDelimiter\ceil{\lceil}{\rceil}

\DeclareMathOperator{\diag}{diag}

\begin{acronym}
\acro{AWGN}{additive white Gaussian noise}
\acro{AoA}{angle of arrival}
\acro{BF}{beamformer}
\acro{BS}{base station}
\acro{CP}{cyclic prefix}
\acro{DoA}{direction of arrival}
\acro{DoD}{direction of departure}
\acro{EIRP}{effective isotropic radiated power}
\acro{ELP}{equivalent low-pass}
\acro{FFT}{fast Fourier transform}
\acro{IFFT}{inverse fast Fourier transform}
\acro{i.i.d.}{independent, identically distributed}
\acro{JSC}{joint sensing and communication}
\acro{LOS}{line-of-sight}
\acro{LTE}{long term evolution}
\acro{MCL}{maximum coupling loss}
\acro{MPL}{maximum path loss}
\acro{MIL}{maximum isotropic loss}
\acro{MC}{Monte Carlo}
\acro{MIMO}{multiple-input multiple-output}
\acro{MUSIC}{MUltiple SIgnal Classification}
\acro{MDL}{minimum description length}
\acro{NR}{new radio}
\acro{OFDM}{orthogonal frequency division multiplexing}
\acro{OSPA}{optimal sub-pattern assignment}
\acro{PSD}{power spectral density}
\acro{QPSK}{quadrature phase shift keying}
\acro{RCS}{radar cross-section}
\acro{RF}{radio frequency}
\acro{RMSE}{root mean squared error}
\acro{r.v.}{random variable}
\acro{SSIR}{signal-to-self interference ratio}
\acro{SI}{self interference}
\acro{SNR}{signal-to-noise ratio}
\acro{SU}{single-user}
\acro{SCM}{sample covariance matrix}
\acro{UE}{user equipment}
\acro{ULA}{uniform linear array}
\end{acronym}

\begin{document}

\title{System-Level Analysis of Joint Sensing and Communication based on 5G New Radio}

\author{Lorenzo~Pucci,~\IEEEmembership{Student~Member,~IEEE,}
Enrico~Paolini,~\IEEEmembership{Senior~Member,~IEEE,}
and~Andrea~Giorgetti,~\IEEEmembership{Senior~Member,~IEEE}
\thanks{Part of this work has been presented at the Workshop on Advances in Network Localization and Navigation (ANLN), IEEE Globecom 2021 \cite{PucMatPaoXuGio:C21}.}
\thanks{This work has been carried out in the framework of the CNIT National Laboratory WiLab and the WiLab-Huawei Joint Innovation Center.}
\thanks{The authors are with the Department of Electrical, Electronic, and Information Engineering ``Guglielmo Marconi'' and CNIT/WiLab, University of Bologna, Italy (e-mail: \{lorenzo.pucci3,e.paolini,andrea.giorgetti\}@unibo.it).}
\thanks{\copyright{ 2022 IEEE. Personal use of this material is permitted. Permission from IEEE must be obtained for all other uses, in any current or future media, including reprinting/republishing this material for advertising or promotional purposes, creating new collective works, for resale or redistribution to servers or lists, or reuse of any copyrighted component of this work in other works.}}
}
\maketitle
\IEEEpubidadjcol
\markboth{Accepted for publication in IEEE Journal on Selected Areas in Communications}%
{Pucci~\MakeLowercase{\textit{et al.}}: System-Level Analysis of Joint Sensing and Communication based on 5G New Radio}
\begin{abstract}
This work investigates a multibeam system for \ac{JSC} based on \ac{MIMO} 5G \ac{NR} waveforms. In particular, we consider a \ac{BS} acting as a monostatic sensor that estimates the range, speed, and \ac{DoA} of multiple targets via beam scanning using a fraction of the transmitted power. The target position is then obtained via range and \ac{DoA} estimation. We derive the sensing performance in terms of probability of detection and root mean squared error (RMSE) of position and velocity estimation of a target under \ac{LOS} conditions. Furthermore, we evaluate the system performance when multiple targets are present, using the \ac{OSPA} metric.
Finally, we provide an in-depth investigation of the dominant factors that affect performance, including the fraction of power reserved for sensing.
\end{abstract}
\acresetall


\section{Introduction}
\IEEEPARstart{T}{he} widespread diffusion of increasingly broadband mobile radio networks and the conquest of mmWave by 5G communication systems has exacerbated the spectrum congestion. On the other hand, sensing capabilities via \ac{RF} signals are gaining increasing interest for several applications, such as autonomous driving, assisted living, security, and human-machine interface \cite{ChiGioPao:J18}. In the framework of this trend, we are witnessing an increasing demand for systems exhibiting both sensing and communications capabilities, i.e., systems where radar and communication functionalities share the hardware platform, as well as the frequency band \cite{PauChiBli,Hanzo2020StateoftheArt}. This approach is also known as \ac{JSC} and consists in exploiting the waveforms transmitted by a communication network to perform sensing functions. In this scenario, \ac{OFDM} based signals are considered good candidates both for active and passive radar purposes. For example, in \cite{Barneto} and \cite{FullDuplex} \ac{OFDM}-based communication systems such as \ac{LTE} and 5G \ac{NR} are suggested. In \cite{EveJack} the self-ambiguity function (SAF) and cross-ambiguity function (CAF) for the frequency division duplex (FDD) \ac{LTE} downlink waveform are evaluated. Another possible solution, widely explored in literature, is represented by passive radars relying on signals of opportunity. In this case, the problem of the spectrum sharing is solved by removing the radar sources, i.e., by performing localization and tracking of targets without the need for radar signals emissions but only by exploiting illuminators of opportunity already present in the environment \cite{BarConWin,BerDemHecWilZho}. However, the radar system does not have full knowledge about the transmitted signals and can only perform detection and estimation based on the power, angles, and Doppler information extracted from the received echo \cite{Rahetal:J20}.

In \cite{Hanzo2020StateoftheArt} applications of \ac{JSC} systems are discussed. According to \cite{zhang2021enabling}, \ac{JSC} can be studied focusing on both simple point-to-point communications such as vehicular networks \cite{zhang2018multibeam,Kumetal:J18,DanYehHea:J18,Luoetal:J19,Luoetal:J20,Petetal:J19} (thus finding great applications in autonomous driving) and complex mobile/cellular networks \cite{Rahetal:J20,Liuetal:J18,Liuetal:J18b,Gutetal:C19}, which can potentially revolutionize the current communication-only mobile networks. \ac{JSC} also has the potential of integrating radio sensing into large-scale mobile networks, generating the so-called perceptive mobile networks \cite{Zhaetal:C17,Rahetal:J20,Zhaetal:C17c,Rahetal:C17,Rahetal:C17b}. Moreover, literature on mmWave \ac{JSC} demonstrates its feasibility and potentials in indoor and vehicle networks \cite{Misetal:J19,zhang2018multibeam,Zhaetal:C17b,Luoetal:J19,AllHua:J19,LiuMas:C19,KumVorHea:J20,Doketal:J19}.
In particular, in-depth signal processing aspects of mmWave-based \ac{JSC} with an emphasis on waveform design are provided by \cite{Misetal:J19}.
At mmWave frequencies, \ac{MIMO} systems enable very high capacity links and reduced latencies to the communication users through spatial multiplexing, while also providing augmented sensing capabilities due to accurate \ac{DoA} estimation \cite{5GScenarios,barneto2021multiuser}. 

Some research studies on \ac{JSC} mainly focus on single beam approaches per phased array, having the sensing beam in the same direction of the communication beam \cite{Heath2018SingleBeam}. To overcome this problem, recent studies have focused on using separate coexisting beams for communication and sensing due to the very different requirements of the two functionalities. For example, in \cite{zhang2018multibeam,barneto2020multibeam} a multibeam framework consisting of beamforming design and optimization are proposed to simultaneously allow a steady communication beam towards the \ac{UE} and a sensing beam to scan the environment.

Most of the research efforts on \ac{JSC} have so far been devoted to the design of signal processing techniques aimed at extracting features from the environment, such as the position and speed of a target (for example, a car or a human being) or at inferring the environment itself, such as the mapping or imaging of a room. However, to the authors' knowledge, very few works have investigated the performance of a \ac{JSC} system, especially from the sensing perspective, and provided results in terms of target parameters estimation accuracy with current technology. For this reason, this work aims to address the analysis of a multibeam system for \ac{JSC} based on 5G \ac{NR} to understand the key aspects and their role in governing performance. In particular, the main contributions are the following:

\begin{itemize}
\item We provide an analysis of the performance of a \ac{BS} acting as a sensor in a monostatic configuration that estimates the range, speed, and \ac{DoA} of multiple targets, through numerical simulations. In particular, we provide a detailed analysis of how the system performance is affected by the portion of the total radiated power used for sensing. Furthermore, we propose an algorithm to remove phantom targets that appear because of the sensing method, which relies on beam-scanning impaired by beam sidelobes. 
\item We analyze the \ac{RMSE} of position estimate, obtained by target localization via range and \ac{DoA} estimation, and the accuracy of radial speed estimation for the single-target scenario.
\item We identify the main dominant factors affecting performance and compare two system setups operating at sub-$6\,$GHz and mmWave frequencies.
\item We use the \ac{OSPA} metric to study the performance of the considered system for the multi-target scenario at mmWave frequencies in terms of localization and detection capabilities.
\end{itemize}

Throughout this paper, capital boldface letters denote matrices, lowercase bold letters indicate vectors, $(\cdot)^T$, $(\cdot)^\dag$, and $(\cdot)^c$ stand for transpose, conjugate transpose, and conjugate of a vector/matrix respectively, $\bigl\| \cdot \bigr\|_p$ is the $p$-norm operator, and $\ceil*{\cdot}$ is the ceiling function. 

The paper is organized as follows. In Section~\ref{sec:intro}, the system model and the proposed \ac{JSC} scheme are described. Section~\ref{sec:estim} presents the adopted estimation techniques and the repeated targets pruning procedure. Section~\ref{sec:ospa} discusses the use of the \ac{OSPA} metric to evaluate the performance of the multi-target system. In Section~\ref{sec:sysanalysis}, an extensive performance analysis is presented. We then conclude this article with our remarks in Section~\ref{sec:conclu}.

\section{System Model}\label{sec:intro}
\begin{figure*}[t]
\captionsetup{font=footnotesize,labelfont=footnotesize}
    \centering
    \includegraphics[width=0.86\textwidth]{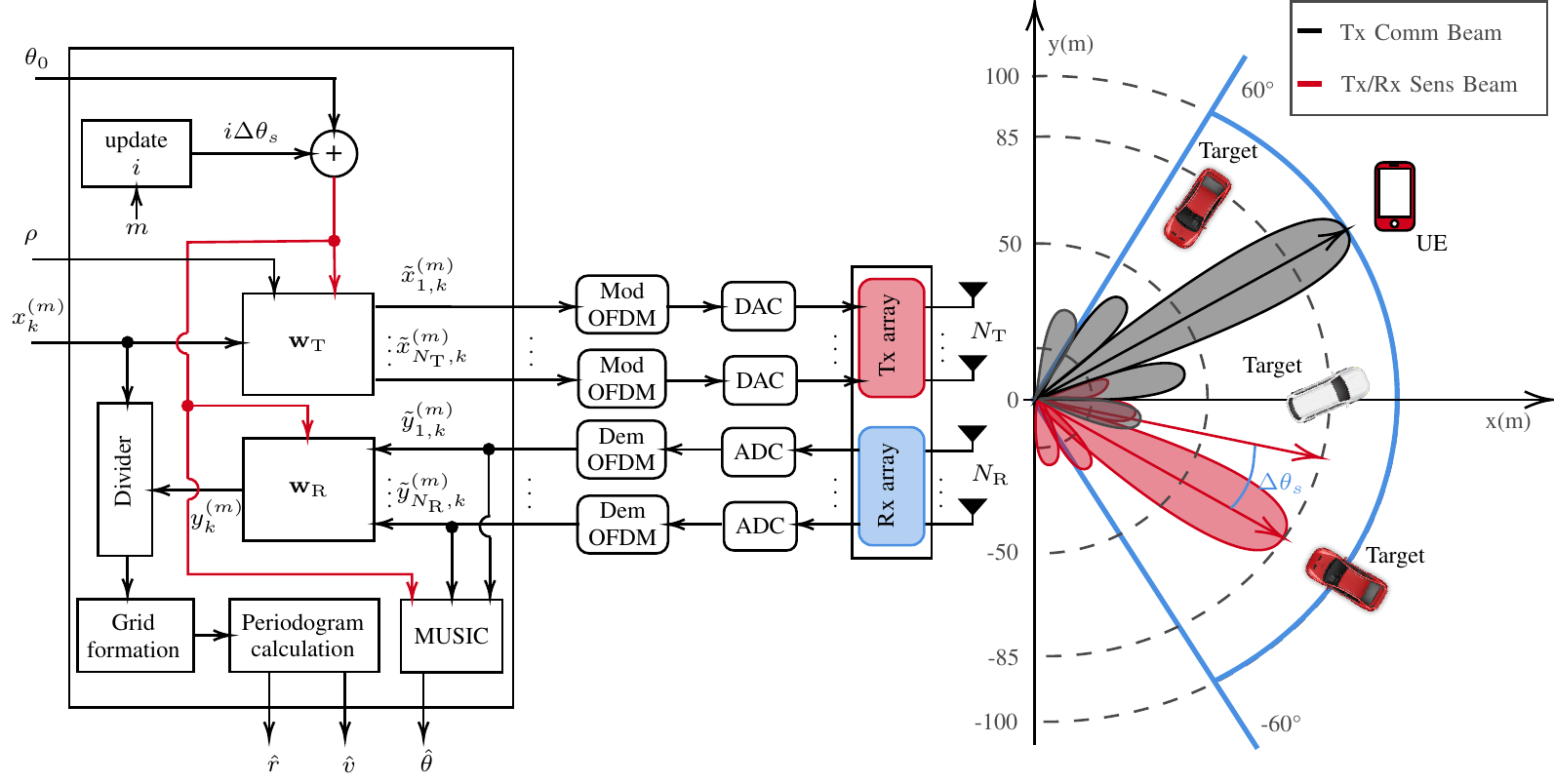}
    \caption{Block diagram of the 5G NR-based sensor with multibeam capability for joint communication and sensing.}
    \label{fig:schematic}
\end{figure*}

As depicted in Fig.~\ref{fig:schematic}, a multiple antennas \ac{OFDM} system is considered. The \ac{JSC} system consists of a transmitter (Tx) antenna array with $N_\mathrm{T}$ elements and of an receiver (Rx) antenna array with $N_\mathrm{R}$ elements, used for communication and sensing, respectively. For both Tx and Rx we assume a \ac{ULA} with half-wavelength separation, i.e., $d = \lambda/2$ with $\lambda=c/f_\mathrm{c}$, $c$ the speed of light, and $f_\mathrm{c}$ the carrier frequency.
The communication system transmits a 5G \ac{NR} waveform with $M$ \ac{OFDM} symbols and $K$ active subcarriers to a \ac{UE} in the cell \cite{asplund2020advanced}.\footnote{Without loss of generality, we consider one user; however, during the scan period described in Section~\ref{sec:multibeam}, the \ac{UE} may change according to the multiple access rule established for communication.} The \ac{ELP} representation of the signal transmitted by the $n$th antenna can be written as
\begin{equation}
       \label{eq:base-band}
        s_n(t) = \sum_{m=0}^{M-1}\left( \sum_{k=0}^{K-1}\tilde{x}_{n,k}^{(m)}e^{j2 \pi \frac{k}{T}t}\right)g(t-mT_\mathrm{s})  
\end{equation}
where $\tilde{x}_{n,k}^{(m)}$ is the modulation symbol, taken from a complex modulation alphabet, to be transmitted to the \ac{UE} at the $m\text{th}$ \ac{OFDM} symbol and $k\text{th}$ subcarrier, mapped through digital precoding at the $n\text{th}$ transmitting antenna, $g(t)$ is the employed pulse, $\Delta f = 1/T$ is the subcarrier spacing, and $T_\mathrm{s}$ is the \ac{OFDM} symbol duration including the \ac{CP}. 

\subsection{Joint waveform}
The vector $\tilde{\mathbf{x}}_k^{(m)} \in \mathbb{C}^{N_\mathrm{T} \times 1}$ is defined as $\tilde{\mathbf{x}}_k^{(m)} = \mathbf{w}_\mathrm{T} x_k^{(m)}$, where $\mathbf{w}_\mathrm{T} \in \mathbb{C}^{N_\mathrm{T} \times 1}$ is the precoder vector used to map each modulation symbol, $x_k^{(m)}$, to the transmitting antennas. In particular, we consider a multibeam system where the power of the \ac{OFDM} signal to be transmitted is split between communication and sensing, namely, the total available power is in part exploited to sense the environment and in part directed to the \ac{UE} \cite{zhang2018multibeam,barneto2020multibeam}.  
Therefore, the transmitting \ac{BF} vector $\mathbf{w}_\mathrm{T}$ can be written as \cite{zhang2018multibeam}
\begin{equation}
\mathbf{w}_\mathrm{T} = \sqrt{\rho}\,\mathbf{w}_\mathrm{T,s} + \sqrt{1-\rho}\,\mathbf{w}_\mathrm{T,c}
\label{eq:BFvector}
\end{equation}
where $\rho \in [0,1]$ is the parameter used to control the fraction of the total power apportioned to the two directions, while $\mathbf{w}_\mathrm{T,c}$ and $\mathbf{w}_\mathrm{T,s}$ are the communication and the sensing \ac{BF} vectors, respectively.
The latter are defined as \footnote{Without loss of generality, we perform a beam steering operation adopting a multibeam approach for both sensing and communication. Other methods exist in the literature for beamforming, e.g., based on optimization techniques, further improving performance \cite{Luoetal:J19,Luoetal:J20}.} \cite{asplund2020advanced}
\begin{equation}
\mathbf{w}_\mathrm{T,c} = \frac{\sqrt{P_\mathrm{T} G_\mathrm{T}^\mathrm{a}}}{N_\mathrm{T}}\,\mathbf{a}_\mathrm{T}^{c}(\theta_\mathrm{T,c}) \quad
\end{equation}
\begin{equation}
\mathbf{w}_\mathrm{T,s} = \frac{\sqrt{P_\mathrm{T} G_\mathrm{T}^\mathrm{a}}}{N_\mathrm{T}}\,\mathbf{a}_\mathrm{T}^{c}(\theta_\mathrm{T,s})
\label{eq:txBF}
\end{equation}
where $G_\mathrm{T}^\mathrm{a}$ is the transmit array gain along the beam steering direction (where such a gain is maximum),  $P_\mathrm{T}G_\mathrm{T}^\mathrm{a}$ is the \ac{EIRP}, $\mathbf{a}_\mathrm{T}(\theta_\mathrm{T,c}) \in \mathbb{C}^{N_\mathrm{T} \times 1}$ and $\mathbf{a}_\mathrm{T}(\theta_\mathrm{T,s}) \in \mathbb{C}^{N_\mathrm{T} \times 1}$ are the steering vectors for communication and sensing, respectively. The spatial steering vector for a \ac{ULA} at a given \ac{DoA}/\ac{DoD} $\theta_l$ is  \cite[Chapter~9]{Richards},\cite[Chapter 5]{asplund2020advanced}
\begin{equation}
\mathbf{a}(\theta_l) = \left [1, e^{j2\pi d \sin(\theta_l)/\lambda}, \dots, e^{j 2\pi (N_\mathrm{a}-1) d \sin(\theta_l)/\lambda} \right]^T 
\label{eq:steering2}
\end{equation}
where $N_\mathrm{a}$ is the number of array antenna elements. Since a half-wavelength separation is considered, the expression \eqref{eq:steering2} reduces to
\begin{equation}
    \mathbf{a}(\theta_l) = [1, e^{j\pi \sin(\theta_l)}, \dots, e^{j\pi (N_\mathrm{a}-1)\sin(\theta_l)}]^T.
\end{equation}

Looking at \eqref{eq:BFvector}, it is evident the trade-off between the performance of the communication and sensing functions. To guarantee certain sensing capabilities, it is necessary to reserve a fraction of the total power available for it, with a consequent reduction in communication coverage.
To study how the communication system coverage changes by varying the \ac{EIRP}, some metrics can be used according to the 3GPP Technical Report in \cite{TR38}. In particular, \ac{MCL}, \ac{MPL} and \ac{MIL} are the metrics used in 5G NR systems to express the coverage of the communication system \cite{TR38,moloudi2021coverage}. These metrics differ for some terms, but they share the main idea of maximum loss that the communication system can tolerate and still be operational. In particular, differently from \ac{MCL}, \ac{MIL} and \ac{MPL} include also the antenna gains. Moreover, the \ac{MIL} metric takes into account parameters such as shadow fading and penetration margins. A detailed analysis of this metric is out of the scope of this paper, but it is important to highlight that the fraction of power $\rho$ reserved for sensing results in a reduction of \ac{MPL} and \ac{MIL} by a factor $10\text{log}_{10}(\rho)\,$dB.

\subsection{Received signal} \label{sec:rxSignal}
The vector $\mathbf{\tilde{y}}_k^{(m)} \in \mathbb{C}^{N_\mathrm{R} \times 1}$ of the received modulation symbols at each antenna after the \ac{FFT} block in the \ac{OFDM} receiver, is given by
\begin{equation}
    \tilde{\mathbf{y}}_k^{(m)} = \mathbf{H}_k^{(m)} \tilde{\mathbf{x}}_k^{(m)} + \tilde{\boldsymbol{\nu}}_k^{(m)} + \tilde{\mathbf{n}}_k 
    \label{eq:y_tilde}
\end{equation}
where $ \mathbf{H}_k^{(m)} \in \mathbb{C}^{N_\mathrm{R} \times  N_\mathrm{T}}$ is the channel matrix for the $m$th \ac{OFDM} symbol and the $k$th subcarrier, $\tilde{\boldsymbol{\nu}}_k^{(m)} \in \mathbb{C}^{N_\mathrm{R} \times 1}$ is the vector whose elements represent the \ac{SI} due to imperfect Tx–Rx isolation at each receiving antenna, and $\tilde{\mathbf{n}}_k  \in \mathbb{C}^{N_\mathrm{R} \times 1}$ is the \ac{AWGN} vector whose entries are \ac{i.i.d.} \acp{r.v.}, having circularly symmetric zero mean Gaussian distribution with variance $\sigma_\mathrm{N}^2$.

Considering $L$ point target reflections, the channel matrix can be written as
\begin{equation}\label{eq:channel-matrix}
    \mathbf{H}_k^{(m)} = \sum_{l = 1}^{L} \underbrace{\alpha_l e^{j2\pi m T_\mathrm{s} f_{\mathrm{D},l}}e^{-j2\pi k \Delta f \tau_l}}_{\triangleq \beta_l} \mathbf{a}_\mathrm{R}(\theta_l)\mathbf{a}^T_\mathrm{T}(\theta_l)
\end{equation}
where $\tau_l$, $f_{\mathrm{D},l}$, and $\theta_l$ are the round-trip delay, the Doppler shift, and the \ac{DoA} of the $l$th target, respectively. The term $\alpha_l = \left|\alpha_l\right|e^{j\phi_l}$ is the complex amplitude which includes phase shift and attenuation along the $l$th propagation path. 
The array response vector at the receiver for sensing is denoted in \eqref{eq:channel-matrix} by $\mathbf{a}_\mathrm{R}(\theta_l)$. 
To simplify the presentation of the \ac{DoA} estimation method, \eqref{eq:channel-matrix} can be recast in the more compact form
\begin{equation}
 \mathbf{H}_k^{(m)} = \mathbf{A}_\mathrm{R}(\boldsymbol{\theta})\boldsymbol{\Sigma}\mathbf{A}_\mathrm{T}^{T}(\boldsymbol{\theta})
 \label{eq:channel-matrix-matform}
\end{equation}
where $\mathbf{A}_\mathrm{R}(\boldsymbol{\theta})=[\mathbf{a}_\mathrm{R}(\theta_1),\dots, \mathbf{a}_\mathrm{R}(\theta_L)] \in \mathbb{C}^{N_\mathrm{R} \times L}$ and $\mathbf{A}_\mathrm{T}(\boldsymbol{\theta})=[\mathbf{a}_\mathrm{T}(\theta_1),\dots, \mathbf{a}_\mathrm{T}(\theta_L)] \in \mathbb{C}^{N_\mathrm{T} \times L}$ are the steering matrices for the targets' directions $\boldsymbol{\theta} = [\theta_1, \theta_2,\dots, \theta_L]$, and $\boldsymbol{\Sigma} = \diag(\beta_1, \beta_2,\dots, \beta_L) \in \mathbb{C}^{L \times L}$  is the diagonal matrix of the channel coefficients. 

For what concerns the \ac{SI} term, $\tilde{\boldsymbol{\nu}}_k^{(m)}$, in \eqref{eq:y_tilde}, each element $n$ of this vector can be considered as the signal scattered by a static target with an almost null distance from the receiver, i.e., with $f_\mathrm{D,SI} = 0$ and $\tau_\mathrm{SI} = 0$, thus it can be written as $\tilde{\nu}_{n,k}^{(m)} = \alpha_{\mathrm{SI},n} x_k^{(m)}$, where $\alpha_{\mathrm{SI},n} = \left|\alpha_{\mathrm{SI},n}\right|e^{j\phi_{\mathrm{SI},n}}$ is the complex amplitude, which includes phase shift and attenuation of the \ac{SI} contribution at the $n$th receiving antenna element \cite{Zeng2018, FullDuplex}. As for the targets, all the attenuation factors $\alpha_{\mathrm{SI},n}$ are assumed to be the same for all the receiving antennas. Therefore, the \ac{SSIR} for the eco generated by the target $l$ at each receiving antenna is given by
\begin{equation}
    \mathrm{SSIR} = \frac{|\alpha_l|^2}{|\alpha_\mathrm{SI}|^2}.
\end{equation}

\indent Starting from \eqref{eq:y_tilde}, by performing spatial combining through the receiving \ac{BF} vector, $\mathbf{w}_\mathrm{R} = \mathbf{a}_\mathrm{R}^c(\theta_\mathrm{R,s})$, we have the received symbol $y_k^{(m)} = \mathbf{w}_\mathrm{R}^T \tilde{\mathbf{y}}_k^{(m)}$,
which, using \eqref{eq:channel-matrix-matform}, can be expressed as
\begin{equation}\label{eq:receivedsymb}
    y_k^{(m)} = \mathbf{w}_\mathrm{R}^T \mathbf{A}_\mathrm{R}(\boldsymbol{\theta})\boldsymbol{\Sigma}\mathbf{A}_\mathrm{T}^{T}(\boldsymbol{\theta})\tilde{\mathbf{x}}_k^{(m)} + \mathbf{w}_\mathrm{R}^T \tilde{\boldsymbol{\nu}}_k^{(m)} + \mathbf{w}_\mathrm{R}^T \tilde{\mathbf{n}}_k.
\end{equation}

\subsection{Beam-scanning}\label{sec:multibeam}
As mentioned above, the considered system is a multibeam \ac{JSC} scheme, with a beam pointing to the \ac{UE} and a beam pointing sequentially to different directions to sense the environment.

Referring to Fig.~\ref{fig:schematic}, during a scan the \ac{DoD} and the \ac{DoA} for sensing are the same. Specifically, we have
\begin{equation}
    \theta_\mathrm{T,s} = \theta_\mathrm{R,s} = \theta_0 + i\,\Delta\theta_\mathrm{s}  \qquad i=0,\dots,N_\mathrm{dir}-1
\end{equation}
where $\theta_0$ is the starting scan direction, $\Delta \theta_\mathrm{s} $ is the scan angle step, $i$ is the index used to update the direction, and $N_\mathrm{dir}$ is the number of directions explored to perform a complete scan from $-\theta_0$ to $\theta_0$.
For each sensing direction, a number of \ac{OFDM} symbols $M_\mathrm{s}<M$ is acquired from the receiver system. Therefore, since a 5G \ac{NR} frame with $M$ symbols lasts $T_\mathrm{f} = 10$ ms, by fixing $N_\mathrm{dir}$ it is possible to determine the number of frames and the time required to complete a scan as:
\begin{equation}
        N_\mathrm{f} = \ceil*{\frac{M_\mathrm{s}N_\mathrm{dir}}{M}}, \qquad
    T_\mathrm{scan} = T_\mathrm{f} N_\mathrm{f}.
    \label{eq:Nf}
\end{equation}

The \ac{OFDM} symbols collected in each direction are used to estimate range, Doppler and \ac{DoA} of the target. 

\subsection{Sensor-target-sensor path} \label{sec:SNR}
In \ac{LOS} propagation conditions the power received at a given array element from the $l$th path, illuminated by the sensing beam, is proportional to $|\alpha_l|^2$ and given by \cite{Richards}
\begin{equation}
    P_{\mathrm{R},l} =  \rho \cdot  \frac{P_\mathrm{T} G_\mathrm{T}^\mathrm{a} G_\mathrm{R} c^2 \sigma_{\mathrm{RCS},l}}{(4\pi)^3 f_\mathrm{c}^2 d_l^4}\cdot \gamma_l
    \label{eq:rx-power}
 \end{equation} 
where $\sigma_{\mathrm{RCS},l}$ is the \ac{RCS} of the  point target $l$,  $d_l$ is the distance between the $l$th target and the \ac{BS}, $G_\mathrm{R}$ is the single element antenna gain at RX, and $\gamma_l=|\mathrm{AF}(\theta_\mathrm{T,s}-\theta_l)|^2 \in [0,1]$ where $\mathrm{AF}(\theta)$ is the normalized array factor at Tx that considers the non-perfect alignment between the target \ac{DoA} and the sensing direction \cite{Orf:B16}; when $\theta_l=\theta_\mathrm{T,s}$ then $\gamma_l=1$.
The \ac{SNR} at the single receiving antenna element related to the $l$th target is defined as 
\begin{equation}
    \text{SNR}_l = \frac{P_{\mathrm{R},l}}{N_0 K \Delta f}
    \label{eq:SNR-path-loss}
\end{equation}
where $P_{\mathrm{R},l}$, the received power from the $l$th path, is given in \eqref{eq:rx-power}, and $N_0$ is the one-sided noise \ac{PSD} at each antenna element. When convenient, by normalizing the received symbols  after  the  \ac{FFT} in the \ac{OFDM} receiver as $\mathbb{E}\{\abs{\tilde{y}_{n,k}^{(m)}}^2\} = 1$, \eqref{eq:SNR-path-loss} reduces to $\text{SNR}_l = 1/\sigma_\mathrm{N}^2$. 

\section{Estimation of Target Parameters and Detection}\label{sec:estim}
This section introduces \ac{MUSIC} for \ac{DoA} estimation and periodogram-based frequency estimation for range and velocity evaluation. The estimation methods are performed for each sensing beam step in which $M_\mathrm{s}$ \ac{OFDM} symbols are collected. To simplify the notation, we drop the scan index $i$.

\subsection{Estimation of the number of targets and DoAs}
\Ac{DoA} estimation is performed by \ac{MUSIC} that requires knowledge of the noise subspace, which in turn needs the number of targets to be known. Noise subspace can be identified via the covariance matrix of the received vector \eqref{eq:y_tilde} $\mathbf{R} = \mathbb{E}\bigl\{\tilde{\mathbf{y}}_k^{(m)}\tilde{\mathbf{y}}_k^{(m) \dag }\bigr\} \in \mathbb{C}^{N_\mathrm{R}\times N_\mathrm{R}}$. In fact, since the noise is zero mean and independent of the target echoes, it follows that 
the $N_\mathrm{R}-L$ smallest eigenvalues of $\mathbf{R}$ are all equal to the noise power $\sigma_\mathrm{N}^2$ and the corresponding eigenvectors identify the noise subspace.\footnote{As required by \ac{MUSIC} we consider $L<N_\mathrm{R}$, i.e, the number of targets is less than the number of sensing array elements.}
Since the covariance matrix is not known a priori, the \ac{SCM} can be used instead 
\cite{barneto2021multiuser}. It is given by
\begin{equation}
    \widehat{\mathbf{R}} = \frac{1}{KM_\mathrm{s}} \sum_{m=0}^{M_\mathrm{s}-1} \sum_{k=0}^{K-1} \tilde{\mathbf{y}}_k^{(m)} \tilde{\mathbf{y}}_k^{(m) \dag}.
\end{equation}

The number of sources (target echoes in our scenario) can be estimated by model order selection based on information theoretic criteria \cite{WaxKailath,MarGioChi:J15sp}. 
The approach starts by performing eigenvalue decomposition of the \ac{SCM} of the observed vectors, $\widehat{\mathbf{R}}=\mathbf{U}\boldsymbol{\Lambda}\mathbf{U}^\dag$,
where the columns of $\mathbf{U}\in \mathbb{C}^{N_\mathrm{R}\times N_\mathrm{R}}$ are the eigenvectors and $\boldsymbol{\Lambda}=\mathrm{diag}(\lambda_1,\dots,\lambda_{N_\mathrm{R}})$ is a diagonal matrix with eigenvalues sorted in descending order, i.e., $\lambda_1\geq \lambda_2 \geq \dots \geq \lambda_{N_\mathrm{R}}$. Using the \ac{MDL} criterion, the estimated number of targets (considering that we are illuminating only targets within the sensing beam in the $i$th direction) is 
\begin{equation}\label{eq:mdl}
L_\mathrm{m}=\!\!\operatorname*{arg\,min}_{s\in\{0,\dots,N_\mathrm{R} -1\}} \!\!\{\textrm{MDL}(s)\}
\end{equation}
with
\begin{equation}
\begin{split}
    \textrm{MDL}(s) &= -\ln \left( \frac{\prod_{i=s+1}^{N_\mathrm{R}} \lambda_i^{1/(N_\mathrm{R}-s)}}{\frac{1}{N_\mathrm{R}-s} \sum_{i=s+1}^{N_\mathrm{R}} \lambda_i} \right)^{\!\!\!(N_\mathrm{R} -s)KM_\mathrm{s}}\\
    &+ \frac{1}{2} s(2N_\mathrm{R}-s)\ln(KM_\mathrm{s}).
\end{split}
\end{equation}
The \ac{MUSIC} algorithm then starts from $\widetilde{\mathbf{U}}\in \mathbb{C}^{N_\mathrm{R}\times (N_\mathrm{R}-L_\mathrm{m})}$, the submatrix containing the $N_\mathrm{R}-L_\mathrm{m}$ eigenvectors corresponding to the smallest eigenvalues, $\lambda_{L_\mathrm{m}+1},\dots,\lambda_{N_\mathrm{R}}$, where such eigenvectors represent a good approximation of the noise subspace. Next, the pseudo-spectrum function, whose peaks reveal the presence of incoming signals, can be obtained as \cite{Sch:J86}
\begin{equation}
    f_\mathrm{m}(\theta)=\frac{1}{\bigl\|\widetilde{\mathbf{U}}^\dag\mathbf{a}(\theta)\bigr\|_2^2}.
    \label{eq:pseudo-spectrum}
\end{equation}
The peak locations in $f_\mathrm{m}(\theta)$ are the \ac{DoA} estimates $\widehat{\theta}$.
However, as it will be better explained in Section~\ref{sec:sysanalysis}, in each sensing direction we search for a local maximum of \eqref{eq:pseudo-spectrum} in a limited angle range $[\theta_\mathrm{min}, \theta_\mathrm{max}]$, which depends on the beamwidth of the array response. The \ac{DoA} estimate in each direction is thus given by
\begin{equation}
   \widehat{\theta} = \operatorname*{arg\,max}_{\theta \in [\theta_\mathrm{min}, \theta_\mathrm{max}]} \{f_\mathrm{m}(\theta)\}. 
  \label{eq:theta_hat}
\end{equation}

\subsection{Detection and range-Doppler estimation}
For the range-Doppler profile evaluation, we start from the received symbols \eqref{eq:receivedsymb} from which, by expanding the matrix multiplications, we obtain
\begin{equation}
    y_k^{(m)} =\left( \sum_{l=1}^L \beta_l \Upsilon(\theta_\mathrm{T,s},\theta_\mathrm{R,s},\theta_l)\right) x_k^{(m)} + n_k
\end{equation}
where $n_k=\mathbf{w}_\mathrm{R}^T \tilde{\mathbf{n}}_k$ and $\Upsilon(\theta_\mathrm{T,s},\theta_\mathrm{R,s},\theta_l)\in \mathbb{C}$ is a factor which accounts for the gain due to the array response vector at Tx and Rx and the \ac{DoA} of the target. Since the range and velocity of targets are embedded in $\beta_l$, first, a division is performed to remove the unwanted data symbols \cite{Braun}, i.e., $g_k^{(m)}= y_k^{(m)}/x_k^{(m)}$, which leads to 
\begin{equation}
    g_k^{(m)}= \sum_{l=1}^L \alpha_l e^{j2\pi m T_\mathrm{s} f_{\mathrm{D},l}}e^{-j2\pi k \Delta f \tau_l} \Upsilon(\theta_\mathrm{T,s},\theta_\mathrm{R,s},\theta_l) + \nu_k
    \label{eq:G}
\end{equation}
where $\nu_k=n_k/x_k^{(m)}$. Note that \eqref{eq:G} contains, for each target, two complex sinusoids whose frequencies are related to $f_{\mathrm{D},l}$ and $\tau_l$, while $\alpha_l$ and $\Upsilon(\theta_\mathrm{T,s},\theta_\mathrm{R,s},\theta_l)$ are constant terms.

Starting from \eqref{eq:G}, a periodogram can be computed in order to estimate range and speed of the target as \cite{Braun,braun2010maximum,FullDuplex}
\begin{equation}\label{eq:period}
    \mathcal{P} (q, p) = \left|\sum_{k=0}^{K_\mathrm{p}-1} \biggl( \sum_{m=0}^{M_\mathrm{p}-1} g_k^{(m)} e^{-j2\pi \frac{mp}{M_\mathrm{p}}}\biggr)e^{j2\pi \frac{kq}{K_\mathrm{p}}}\right|^2
\end{equation}
with $q=0,\dots, K_\mathrm{p}-1$ and $p=0,\dots, M_\mathrm{p}-1$, which consists of $K$ \acp{FFT} of length $M_\mathrm{p}$ and $M_\mathrm{s}$ \acp{IFFT} of length $K_\mathrm{p}$. In this work, $K_\mathrm{p}>K$ is calculated as the next power of two of $K$, whereas $M_\mathrm{p}>M_\mathrm{s}$ is the next power of two of $F_\mathrm{p} \cdot M_\mathrm{s}$, where $F_\mathrm{p}$ is the zero-padding factor to improve speed estimation resolution. 

The periodogram \eqref{eq:period} represents the range-Doppler map from which the first operation performed is target detection by a hypothesis test between $H_0$, where only the noise is present, and $H_1$, which refers to the presence of the target, i.e.,
\begin{equation}\label{eq:hypotest}
    \mathcal{P}(q,p) \overset{H_1}{\underset{H_0}{\gtrless}} \eta.
\end{equation}
The threshold $\eta$ is chosen to ensure a predefined false alarm probability $P_\mathrm{FA}$.
When the sensing beamwidth is relatively small, only one target is likely to be present in a given sensing direction, and if the test \eqref{eq:hypotest} rejects the null hypothesis, it is easy to find the location of the peak in the periodogram
\begin{equation}\label{eq:qp_estimate}
    (\widehat{q},\widehat{p}) = \operatorname*{arg\,max}_{(q,p)} \{\mathcal{P}(q,p)\}
\end{equation}
and evaluate the distance and radial speed of the target as 
\begin{equation}
    \widehat{r} = \frac{\widehat{q}\, c}{2\Delta f K_\mathrm{p}},
\qquad    \widehat{v} = \frac{\widehat{p}\, c}{2 f_\mathrm{c}T_\mathrm{s}M_\mathrm{p}}. \label{eq:rv_estimate}
\end{equation}
The distance and velocity resolutions are intrinsic characteristics of the periodogram and only depend on the 5G \ac{NR} parameters, i.e., number of \ac{OFDM} symbols, number of active subcarriers, subcarrier spacing, and \ac{OFDM} symbol duration, and are given by \cite[Chapter 3]{Braun}
\begin{equation}
    \Delta r = \frac{c}{2\Delta f K_\mathrm{p}},
\qquad     \Delta v = \frac{c}{2 f_\mathrm{c}T_\mathrm{s}M_\mathrm{p}}.
\label{eq:resolution}
\end{equation}
%


\begin{algorithm}[t]
\caption{Pruning redundant target points}
\label{alg:targetPrun}
\begin{algorithmic}[1]
\Require $\mathbf{Z} \leftarrow \mathbf{z}_i =[\widehat{r}_i,\widehat{v}_i,\mathcal{P}(\widehat{r}_i,\widehat{v}_i),\widehat{\theta}_i,f_\mathrm{m}(\widehat{\theta}_i)]$
    \State $\mathbf{Z_\mathrm{sort}} \gets$ sort $\mathbf{Z}$ in decreasing order according to the $3$rd column 
    \State $\mathbf{z}_{\mathrm{prun},1} \gets \mathbf{z}_{\mathrm{sort},1}$ \Comment{copy first row of $\mathbf{Z}_\mathrm{sort}$ in $\mathbf{Z}_\mathrm{prun}$}
    \State $k \gets 1$      \Comment{initialize row index of $\mathbf{Z}_\mathrm{prun}$}
    \For{$i = 2:N_\mathrm{max}$}
        \State $ \mathrm{count} \gets 0$
        \For{$j=1:k$}
            \If{$\widehat{r}_{\mathrm{prun},j} - \epsilon_\mathrm{r} \leq \widehat{r}_{\mathrm{sort},i} \leq \widehat{r}_{\mathrm{prun},j} + \epsilon_\mathrm{r}$ \textbf{and} \\
            \hspace{3em} $\widehat{v}_{\mathrm{prun},j} - \epsilon_\mathrm{v} \leq \widehat{v}_{\mathrm{sort},i} \leq \widehat{v}_{\mathrm{prun},j} + \epsilon_\mathrm{v}$}
            \State $\mathrm{count} \gets \mathrm{count} + 1$
            \State \textbf{break}
            \EndIf
        \EndFor
        \If{$\mathrm{count} = 0$}
            \State $k \gets k + 1$      \Comment{update row index of $\mathbf{Z}_\mathrm{prun}$}
            \State $\mathbf{z}_{\mathrm{prun},k} \gets \mathbf{z}_{\mathrm{sort},i}$
        \EndIf
    \EndFor
    \State $\widehat{L} \gets k$
\end{algorithmic}
\textbf{Output:} $\mathbf{Z}_\mathrm{prun}$ and its number of rows, $\widehat{L}$
\end{algorithm}

\subsection{Pruning redundant target points}\label{sec:targPrun}
As explained above, the considered \ac{JSC} system searches for a peak in the pseudo-spectrum \eqref{eq:pseudo-spectrum} and in the periodogram \eqref{eq:period} for each sensing direction for which the test \eqref{eq:hypotest} chooses the hypothesis $H_1$. 
When a target is detected in a particular direction, it might be detected also in some adjacent directions when the periodogram $\mathcal{P}$ is above threshold because of the beam sidelobes.
These detected points are originated by the same target and are characterized by inaccurate \ac{DoA} estimates. 
As it will be better quantified in Section~\ref{sec:sysanalysis}, this effect is due to the choice of searching the maximum of \ac{MUSIC} pseudo-spectrum in a limited range, as in \eqref{eq:theta_hat}, that reduces the computational cost of searching but may yield multiple detection points per target. To maintain the benefits of local search, we propose a method to thin out redundant target points (hereafter also referred to as repeated targets) that has proven effective.

First, all the collected peaks and estimates are organized in a matrix $\mathbf{Z}$, whose rows are the vectors
\begin{equation}
     \mathbf{z}_i = \left[\widehat{r}_i,\widehat{v}_i,\mathcal{P}(\widehat{r}_i,\widehat{v}_i),\widehat{\theta}_i,f_\mathrm{m}(\widehat{\theta}_i)\right]
     \quad
     i = 1,\dots,N_\mathrm{max}
\end{equation}
where $N_\mathrm{max}\leq N_\mathrm{dir}$ is the number of sensing directions in which the test \eqref{eq:hypotest} rejects the null hypothesis.
Subsequently, these rows are sorted in descending order with respect to the values, $\mathcal{P}(\widehat{r}_i,\widehat{v}_i)$, to form a new matrix $\mathbf{Z}_\mathrm{sort}$. Finally, a check on the elements of $\mathbf{Z}_\mathrm{sort}$ is performed to remove redundant target points, i.e., those with very similar estimates of both distance and radial velocity (within a given range of uncertainty). This results in a new matrix $\mathbf{Z}_\mathrm{prun}$ with a number of rows $\widehat{L} \leq N_\mathrm{max}$.\footnote{Note that $\widehat{L}$ is the estimated number of targets in our approach. This value may differ from $L_\mathrm{m}$ given by \eqref{eq:mdl} because the number of targets detected by \ac{MUSIC} is conditioned on the considered sensing direction. After all, Tx beamforming performs spatial filtering, illuminating predominantly targets within the beamwidth.} The sort operation ensures that only range and speed pairs associated with the largest values of the periodogram are kept between the repeated points. The whole procedure is detailed in Algorithm~\ref{alg:targetPrun}. As it can be seen, the definition of \emph{redundant target point} is linked to the choice of two parameters, $\epsilon_\mathrm{r}$, and $\epsilon_\mathrm{v}$, that account for the measurement uncertainty. In particular, in the algorithm a target indexed with $i$ is considered a repetition of an already detected target denoted with $j$ if its estimated range, $\widehat{r}_i$, and velocity, $\widehat{v}_i$, meet the conditions, $\widehat{r}_j - \epsilon_\mathrm{r} \leq \widehat{r}_i \leq \widehat{r}_j + \epsilon_\mathrm{r}$, and $\widehat{v}_j - \epsilon_\mathrm{v} \leq \widehat{v}_i \leq \widehat{v}_j + \epsilon_\mathrm{v}$, respectively. The choice of the two parameters $\epsilon_\mathrm{r}$ and $\epsilon_\mathrm{v}$ will be discussed in Section~\ref{sec:sysanalysis}. 

\section{Performance Evaluation in the Presence of Multiple Targets}\label{sec:ospa}
This section introduces the performance metric employed to address the concept of miss-distance, or error, in a multi-target system. In particular, when considering a multi-object system, a consistent metric should capture the difference between two sets of vectors (the truth and the estimated), not only in terms of localization error but also in terms of cardinality error. For this reason, in this work, the \ac{OSPA} metric \cite{schuhmacher2008}, \cite{9470927} is used to study the performance of the considered \ac{JSC} system in a multi-target scenario.

The \ac{OSPA} metric is a miss-distance indicator, which summarizes in a unique measure the estimation accuracy in both the number and location of the targets. More precisely, given the true positions of the $L$ targets, $\mathbf{P} = \left[\mathbf{p}_1,\dots,\mathbf{p}_L\right]$, with $\mathbf{p}_l= (x_l,y_l) = (r_l \cos{\theta_l}, r_l \sin{\theta_l})$,\footnote{From now on, and without loss of generality, the monostatic sensor is considered at the origin of a Cartesian coordinate system.}
and the $\widehat{L}$ estimates, $\widehat{\mathbf{P}} = \left[\widehat{\mathbf{p}}_1,\dots,\widehat{\mathbf{p}}_{\widehat{L}}\right]$, the distance between an arbitrary pair of the estimate and the true position, cut off at $\bar{c}>0$, is defined as \cite{schuhmacher2008}
\begin{equation}
    d^{(\bar{c})}\left(\mathbf{p},\widehat{\mathbf{p}}\right) = \min \left\{\bar{c},d\left(\mathbf{p},\widehat{\mathbf{p}}\right)\right\}
\end{equation}
where $d\left(\mathbf{p},\widehat{\mathbf{p}}\right) = \bigl\|\mathbf{p}-\widehat{\mathbf{p}}\bigr\|_2$ is the Euclidean distance between the estimate and the true position, and $\bar{c}$ is the cutoff parameter that determines how the metric penalizes cardinality error with respect to the localization one. Denoting by $\Pi_k$ the set of permutations on $\{1,2,\dots,k\}$ for any $k \in \mathbb{N}$, for $1 \leq q \leq \infty$ and $\bar{c} > 0$, the \ac{OSPA} metric of order $q$ and with cutoff $\bar{c}$ is defined as \cite{schuhmacher2008}
\begin{equation}
\begin{split}
    \bar{d}_q^{(\bar{c})} & \left(\mathbf{P},\widehat{\mathbf{P}}\right) = \\ & \left( \frac{1}{\widehat{L}} \left(\min_{\pi \in \Pi_{\widehat{L}}}
    \sum_{l=1}^{L} \left(d^{(\bar{c})}\left(\mathbf{p}_l,\widehat{\mathbf{p}}_{\pi(l)}\right)\right)^q + \bar{c}^q(\widehat{L}-L)\right)\right)^{1/q}
\end{split}
\end{equation}
if $L \leq \widehat{L}$, and $\bar{d}_q^{(\bar{c})}\left(\mathbf{P},\widehat{\mathbf{P}}\right)=\bar{d}_q^{(\bar{c})}\left(\widehat{\mathbf{P}},\mathbf{P}\right)$ if $L>\widehat{L}$.
Essentially, for $L \leq \widehat{L}$, the \ac{OSPA} distance can be obtained by the following steps:
\begin{enumerate}
    \item Find the $L$-elements subset of $\widehat{P}$ that has the shortest distance to $P$, corresponding to the optimal subset assignment;
    \item If a point $\widehat{\mathbf{p}}_n \in \widehat{\mathbf{P}}$ is not paired with any point in $\mathbf{P}$, let $d_n=\bar{c}$; otherwise, $d_n$ is the minimum value between $\bar{c}$ and the distance between the two points in a pair;
    \item The  \ac{OSPA} distance is given by $\bar{d}_q^{(\bar{c})}\left(\mathbf{P},\widehat{\mathbf{P}}\right) = \left(\left(
    \sum_{l=1}^{L} d_n^q \right)/\widehat{L}\right)^{1/q}$.
\end{enumerate}
The \ac{OSPA} distance can be interpreted as a $q$th order \emph{per-target} error for a multi-object scenario. The metric can be divided into two components, one accounting for localization error and the other for cardinality error. In particular, for $q<\infty$ these components are given by \cite{schuhmacher2008}
\begin{equation}
    \begin{split}
        \bar{e}_{q,\mathrm{loc}}^{(\bar{c})}(\mathbf{P},\widehat{\mathbf{P}}) = &  \left(\frac{1}{\widehat{L}} \min_{\pi \in \Pi_{\widehat{L}}}
    \sum_{l=1}^{L} \left(d^{(\bar{c})}\left(\mathbf{p}_l,\widehat{\mathbf{p}}_{\pi(l)}\right)\right)^q\right)^{1/q}, \\
    \bar{e}_{q,\mathrm{card}}^{(\bar{c})}(\mathbf{P},\widehat{\mathbf{P}}) = & \left(\frac{\bar{c}^q(\widehat{L}-L)}{\widehat{L}} \right)^{1/q}
    \end{split}
\end{equation}
%
\begin{table}[t]
\centering
 \caption{\ac{JSC} system parameters}
 \scalebox{0.8}{%
 \begin{tabular}{l c c}
 \toprule
  5G specification $\rightarrow$ & NR 100 & NR 400\\
\midrule
  $f_\mathrm{c}$ [GHz] & 3.5 & 28 \\ 
$\Delta f$ [kHz] & 30 & 120\\
 Active subcarriers $K$ & 3276 & 3168\\
OFDM symbols per frame $M$ & 280 & 1120\\
OFDM symbols per direction $M_\mathrm{s}$ & 112 & 112\\
   Number of antennas $N_\mathrm{T}=N_\mathrm{R}$  & 10 & \begin{tabular}{c c c} 10 & 50 & 100 \\ \end{tabular}\\
 Array beamwidth $\Delta \Theta$ [°] & 27 & \begin{tabular}{c c c} 27 & 5.3 & 2.6 \\ \end{tabular}\\
 \bottomrule
\end{tabular}
}
\label{NRparam}
\end{table}
\begin{figure*}[t]
\captionsetup{font=footnotesize,labelfont=footnotesize}
    \centering
    \subfloat[][\emph{Angle estimation \ac{RMSE}} \label{fig:RMSE_SI_angle}]
    {\includegraphics[width=0.32\textwidth]{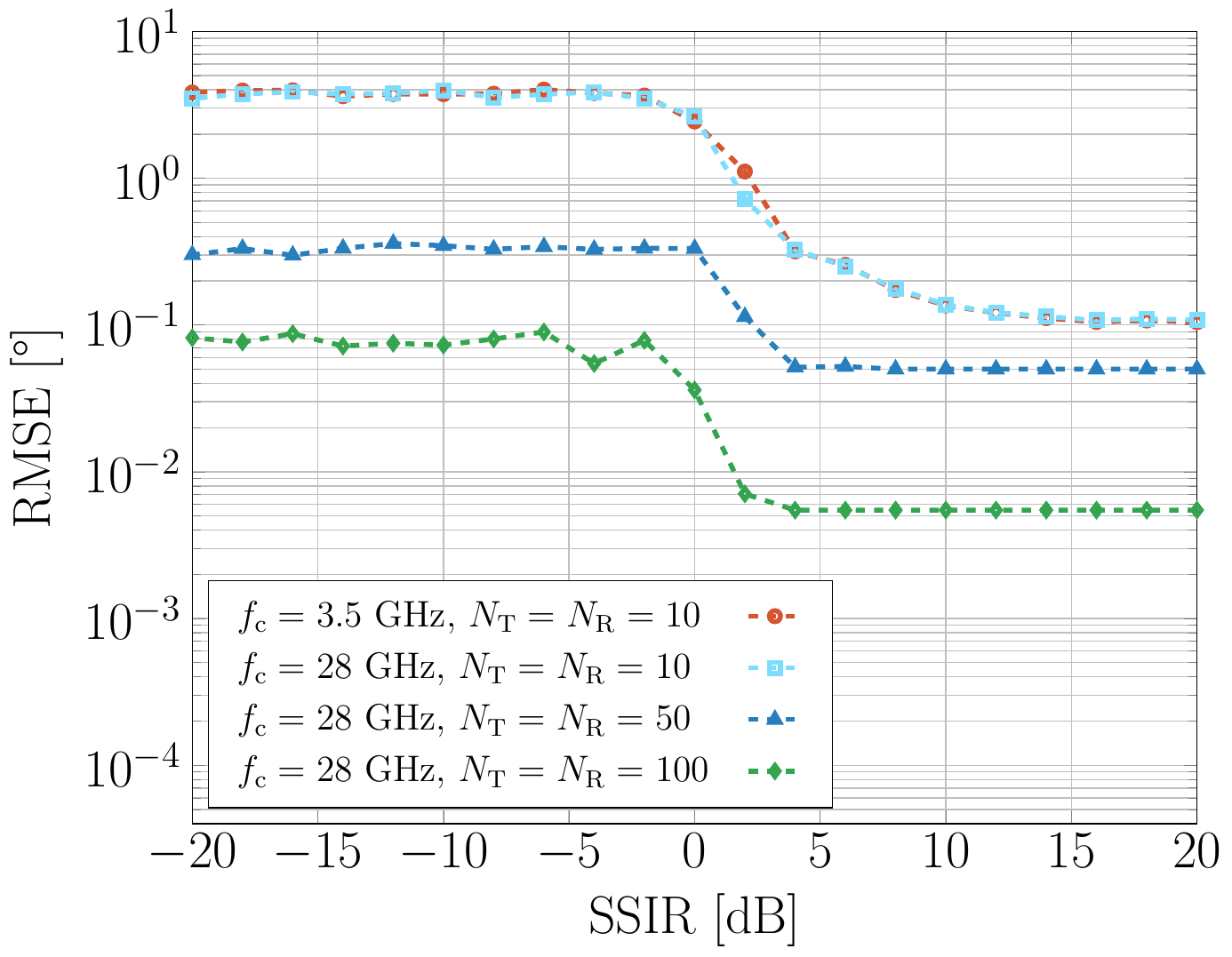}} \quad
    \subfloat[][\emph{Distance estimation \ac{RMSE}} \label{fig:RMSE_SI_dist}]
    {\includegraphics[width=0.32\textwidth]{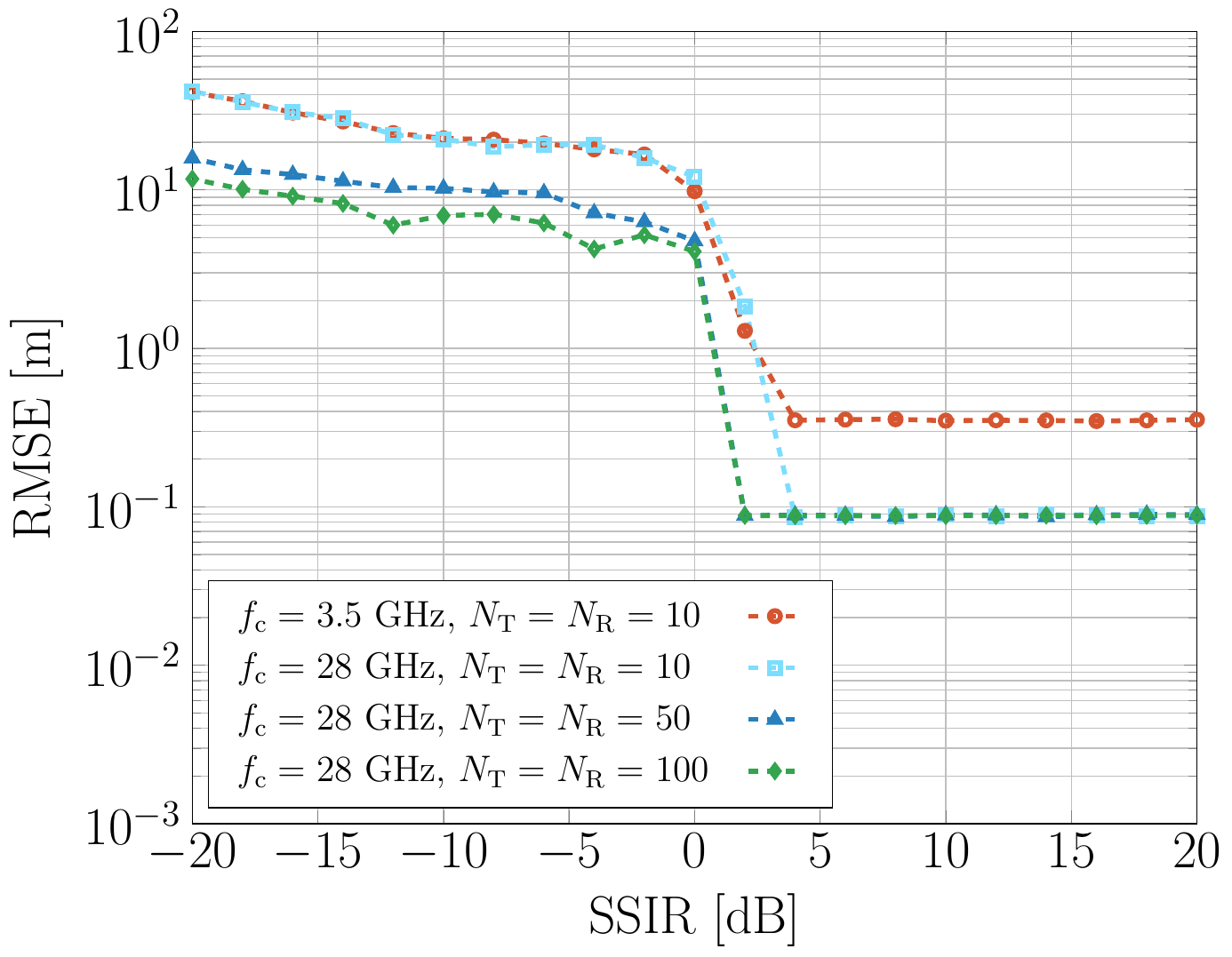}} \quad
    \subfloat[][\emph{Speed estimation \ac{RMSE}} \label{fig:RMSE_SI_speed}]
    {\includegraphics[width=0.32\textwidth]{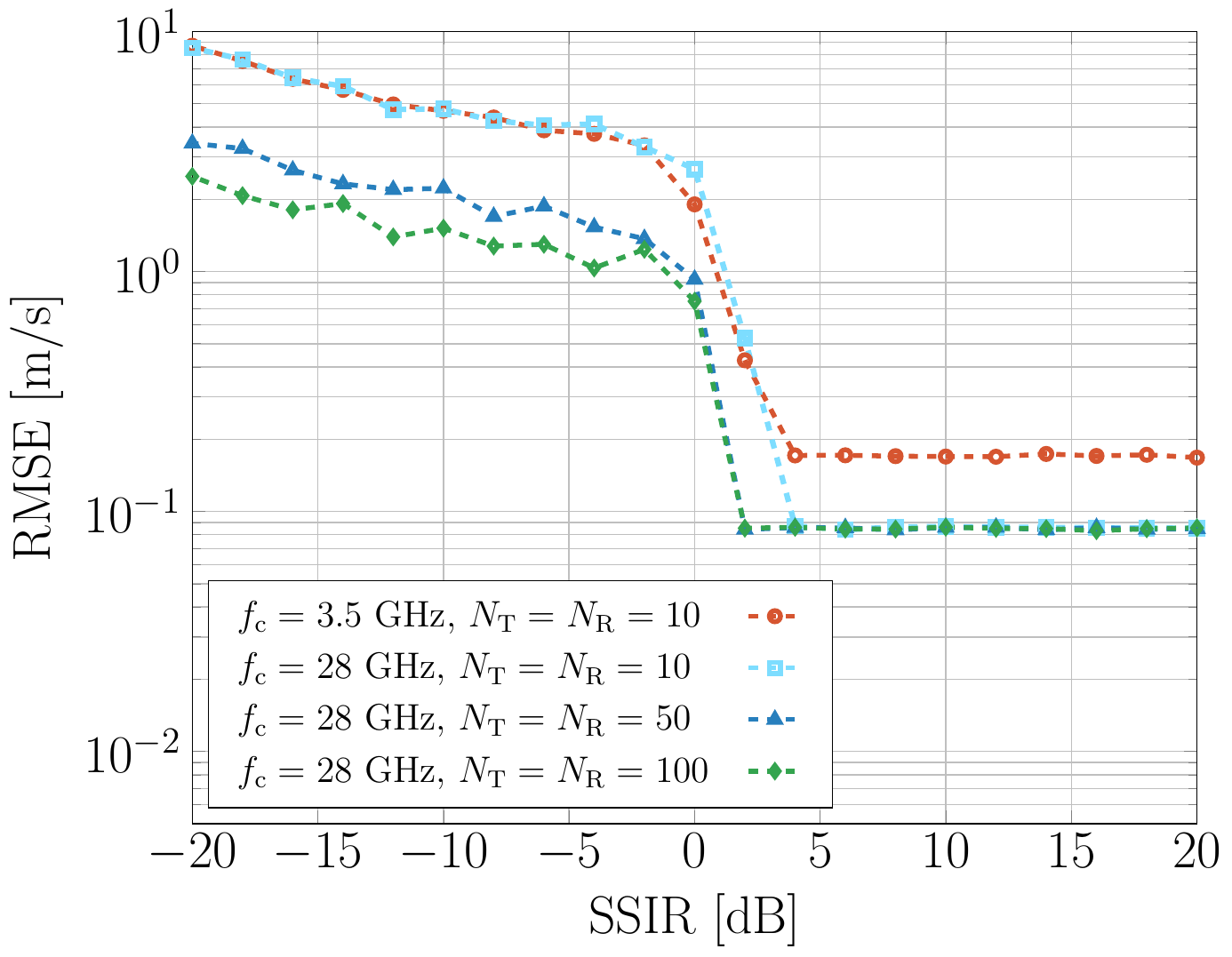}} 
    \caption{Sensing performance as a function of the \ac{SSIR} for \ac{DoA}, distance, and speed estimates, when $\text{SNR} = -20\,$dB.}
    \label{fig:RMSE_SI}
\end{figure*}
if $L\leq \widehat{L}$, and $\bar{e}_{q,\mathrm{loc}}^{(\bar{c})}(\mathbf{P},\widehat{\mathbf{P}}) = \bar{e}_{q,\mathrm{loc}}^{(\bar{c})}(\widehat{\mathbf{P}},\mathbf{P})$,     $\bar{e}_{q,\mathrm{card}}^{(\bar{c})}(\mathbf{P},\widehat{\mathbf{P}}) = \bar{e}_{q,\mathrm{card}}^{(\bar{c})}(\widehat{\mathbf{P}},\mathbf{P})$ if $L>\widehat{L}$. 

In the metric, the value of $q$ determines the sensitivity of the $\bar{d}_q^{(\bar{c})}$ to outlier estimates, while $\bar{c}$ balances the cardinality error component with respect to the localization one, as a part of the total error. As $\bar{c}$ decreases, the localization error becomes dominant compared with the cardinality error, whereas larger values of $\bar{c}$ emphasize the latter. The best choice for $\bar{c}$ to maintain a balance between the two components is any value significantly larger than a typical localization error, but significantly smaller than the maximum distance between objects.

\section{System-Level Analysis}\label{sec:sysanalysis}
System level analysis is carried out through numerical simulations to evaluate the performance of the above-described \ac{JSC} scheme. For all the simulations, 5G \ac{NR} signals compliant with 3GPP Technical Specification in \cite{TS38} are considered. The main 5G \ac{NR} parameters employed for the generation of the standardized signals are summarized in Table~\ref{NRparam}. In addition, a \ac{QPSK} modulation alphabet is used for the generation of the \ac{OFDM} signal.
As it can be seen in Fig.~\ref{fig:schematic}, the considered system scans the environment in the range $[-\theta _0,\theta _0]$, with $\theta_0 = -60$°, and a step $\Delta \theta_\mathrm{s}$. 
The choice of $N_\mathrm{dir}$, and so of $\Delta \theta_\mathrm{s}$, mainly depends on the beamwidth $\Delta \Theta$ of the array response (here referred to $-10\,\text{dB}$ gain with respect to the beam direction) reported in Table~\ref{NRparam}. As expected, when the number of antennas decreases, $\Delta \Theta$ becomes larger, and a lower $N_\mathrm{dir}$ is necessary to avoid blind zones. 
Once $N_\mathrm{dir}$ is chosen, the number of necessary 5G \ac{NR} frames, and consequently, the total time needed to complete a scan cycle, are calculated from \eqref{eq:Nf}.
For each selected direction, the periodogram is obtained from $K$ active subcarriers, which differ between 5G numerologies, and a fixed number of \ac{OFDM} symbols $M_{\mathrm{s}}=112$, with $F_\mathrm{p}=10$, required to perform speed estimation. Furthermore, for what concerns the \ac{DoA} estimation algorithm, the \ac{MUSIC} pseudo-spectrum \eqref{eq:pseudo-spectrum} is computed only in the range $\left[\theta_\mathrm{R,s}-\Delta \Theta/2, \theta_\mathrm{R,s}+\Delta \Theta/2 \right]$, to reduce the processing time and the position error.

As already stated in Section~\ref{sec:intro}, this work addresses the performance analysis of a \ac{JSC} multibeam system considering two different scenarios, single-target, and multi-target. For the former, the primary purpose of the analysis is to derive the \ac{RMSE} of position and speed of the target. When deriving the \ac{RMSE} as a function of the \ac{SNR}, the target is considered aligned with the sensing beam (i.e., $\gamma = 1$) and the noise variance is $\sigma_\mathrm{N}^2 = 1/\text{SNR}$, as mentioned in Section~\ref{sec:SNR}. Whereas when the \ac{RMSE} is evaluated varying the distance of the target, the \ac{SNR} is computed using \eqref{eq:SNR-path-loss} and the following system parameters are considered: the target has an \ac{RCS} equal to $\sigma_\mathrm{RCS}=1\,\text{m}^2$, the \ac{EIRP} is set to $P_\mathrm{T}G_\mathrm{T}^\mathrm{a}=43\,\text{dBm}$, $G_\mathrm{R}=1$, and the noise \ac{PSD} is $N_0=k_\mathrm{B} T_0 F$ where $k_\mathrm{B} = 1.38\cdot 10^{-23}\,\text{JK}^{-1}$ is the Boltzmann constant, $T_0=290\,$K is the reference temperature, and $F=10\,$dB is the receiver noise figure.
The number of \ac{MC} iterations for each \ac{SNR} or distance value is $N_\mathrm{MC}=2000$.
For the multi-object scenario, we consider $L=10$ point targets, one of which is the \ac{UE}, and the same system parameters as the single-target scenario. Two different values of the fraction of power devoted to sensing, $\rho=0.1$ and $\rho=0.3$, a carrier frequency equal to $f_c = 28\,\text{GHz}$, and $N_\mathrm{T}=N_\mathrm{R}=50$ antennas, are considered. In this set of results, the \ac{OSPA} metric, presented in Section~\ref{sec:ospa}, is the performance indicator to summarize the effectiveness of the designed system. The \ac{OSPA} metric is computed for $q=2$, as recommended by \cite{schuhmacher2008}, and $\bar{c}=10\,$m to guarantee a good balance between localization and cardinality error.

\begin{figure*}[t]
\captionsetup{font=footnotesize,labelfont=footnotesize}
    \centering
    \subfloat[][\emph{Angle estimation \ac{RMSE}} \label{figure_a}]
    {\includegraphics[width=0.33\textwidth]{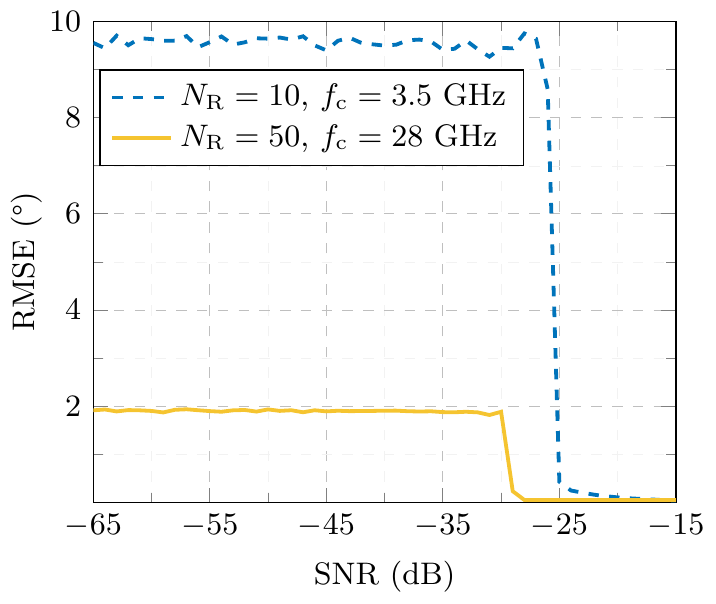}} 
    \subfloat[][\emph{Distance estimation \ac{RMSE}} \label{figure_b}]
    {\includegraphics[width=0.33\textwidth]{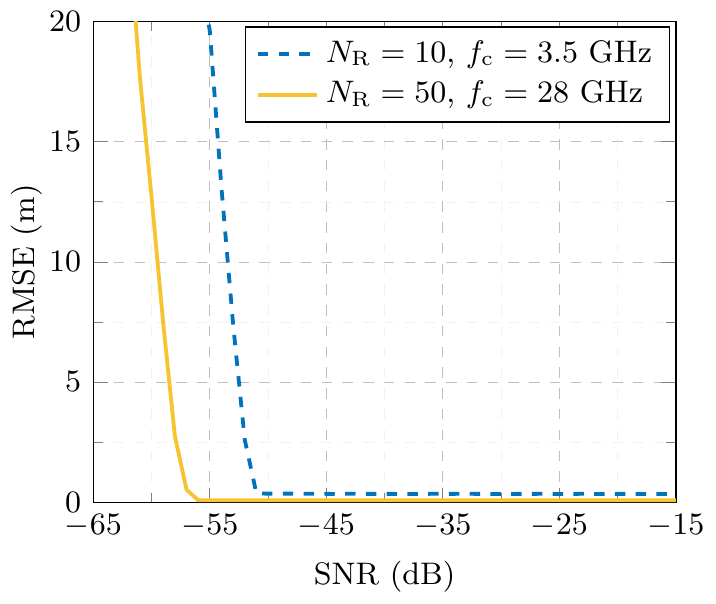}} 
    \subfloat[][\emph{Speed estimation \ac{RMSE}} \label{figure_c}]
    {\includegraphics[width=0.33\textwidth]{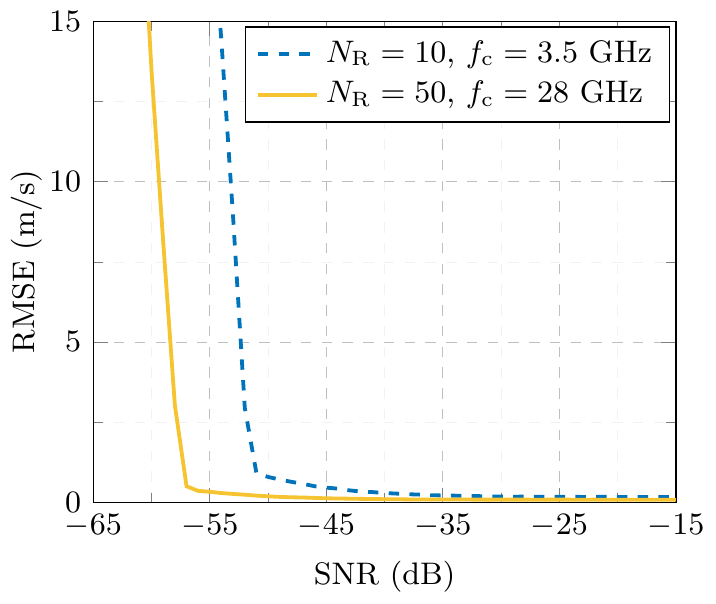}} \\
    \subfloat[][\emph{Normalized position \ac{RMSE}} \label{figure_d}]
    {\includegraphics[width=0.34\textwidth]{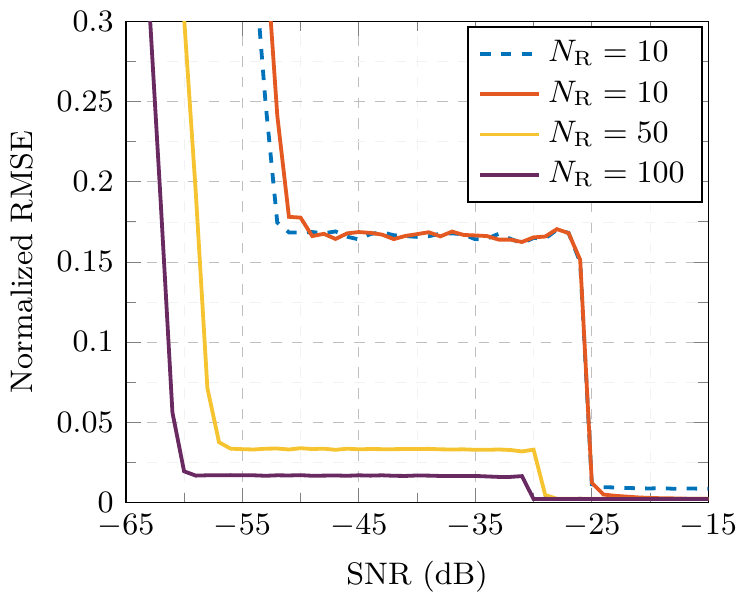}}\qquad
    \subfloat[][\emph{Detection probability} \label{figure_e}]
    {\includegraphics[width=0.34\textwidth]{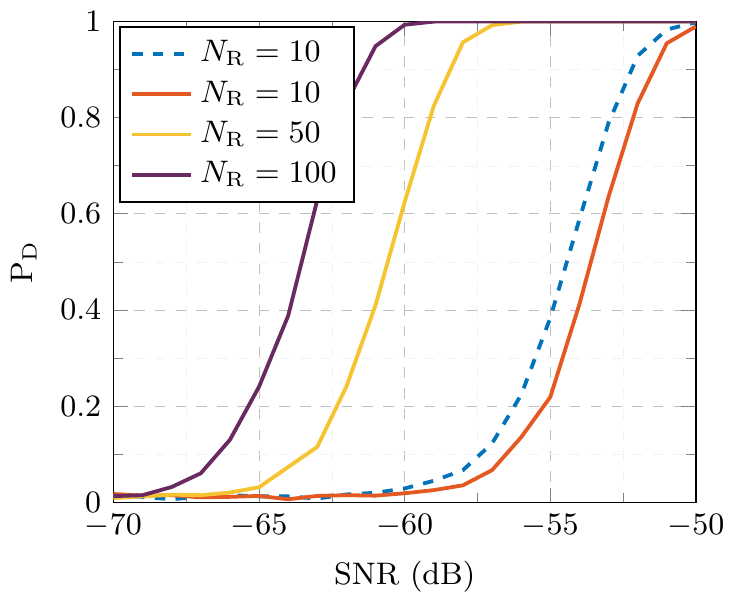}}\\
    \caption{Sensing performance as a function of the \ac{SNR} for distance, \ac{DoA}, speed, and position estimates, and detection probability. The dashed lines represent the results at $f_\mathrm{c}=3.5\,$GHz, whereas the continuous lines represent the results at $f_c=28\,$GHz. In particular, (a), (b) and (c) show the \ac{RMSE} results when the \ac{MIMO} system consists of $N_\mathrm{T}=N_\mathrm{R}=10$ antennas at $f_\mathrm{c}=3.5\,$GHz, and $N_\mathrm{T}=N_\mathrm{R}=50$ antennas at $f_\mathrm{c}=28\,$GHz, whereas (d) and (e) depict the normalized localization error and the detection probability for different number of antennas.}
    \label{fig:RMSE_pos_norm}
\end{figure*}
\begin{figure*}[t]
\captionsetup{font=footnotesize,labelfont=footnotesize}
    \centering
    \subfloat[][\emph{Position \ac{RMSE} at} $f_\mathrm{c} = 3.5\,\text{GHz}$, $N_\mathrm{R}=10$ \label{figure_3a}]
    {\includegraphics[width=0.39\textwidth]{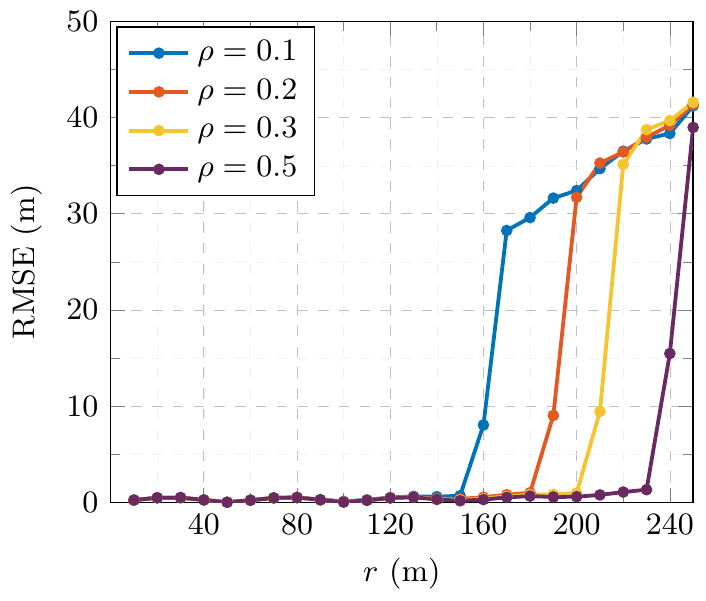}} \qquad \qquad
    \subfloat[][\emph{Position \ac{RMSE} at} $f_\mathrm{c} = 28\,\text{GHz}$, $N_\mathrm{R}=50$ \label{figure_3b}]
    {\includegraphics[width=0.375\textwidth]{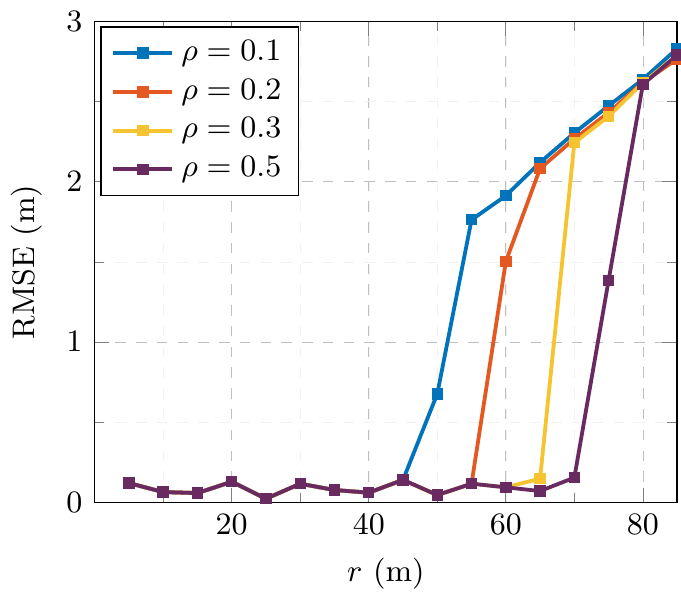}} \\
    \caption{Target localization performance as a function of the sensor-target distance varying the fraction of power $\rho$ reserved for sensing. Note that the maximum range of $250\,$m at $3.5\,$GHz and $85\,$m at $28\,$GHz is within the maximum unambiguous range for the respective numerology detailed in Table~\ref{NRparam} \cite{Braun}.}
    \label{fig:RMSE_rho_Pd}
\end{figure*}
As it will be explained in the following, we will start investigating the impact of \ac{SI} on estimation performance, focusing on the \ac{RMSE} of \ac{DoA}, distance, and speed estimates by varying the \ac{SSIR}  introduced in Section~\ref{sec:rxSignal}.
\subsection{RMSE and detection probability vs SNR and SSIR} \label{sec:RMSEvsSNR}
Let us start analyzing the \ac{SI} issue. Fig.~\ref{fig:RMSE_SI} shows the \ac{RMSE} of \ac{DoA}, range, and radial speed estimates obtained when only a single target is present and $\mathrm{SNR} = -20\,\text{dB}$, by varying the \ac{SSIR} for different 5G \ac{NR} parameters and number of antennas.
As it can be noticed, the system performance quickly degrades for low \ac{SSIR} values, but when $\mathrm{SSIR} \geq 10\,$dB, the \ac{RMSE} of \ac{DoA}, range, and radial speed estimates, reaches a floor where thermal noise is the only limiting factor. Therefore, if proper \ac{SI} suppression can be performed, either through beamforming optimization or through digital cancellation techniques \cite{FullDuplex, barneto2021multiuser}, the estimation error can be kept low. \\ 
\indent From now on, an analysis of the \ac{RMSE} by varying the \ac{SNR} is performed considering \ac{SI} negligible.
In particular, the \ac{SNR} is varied from $-65$ to $-15\,\text{dB}$, while distance, speed, and \ac{DoA} of the target are varied randomly, from one iteration to another, with a uniform distribution from $20$ to $85\,\text{m}$, $-20$ to $20\,\text{m/s}$, and $-60^{\circ}$ to $60^{\circ}$, respectively. In Fig.~\ref{fig:RMSE_pos_norm}, the results obtained for the \ac{RMSE} of distance, angle, speed, and position are shown. As the periodogram used to estimate speed and radial distance of the target is computed on the symbols obtained after spatial combining, the whole estimation process is subject to a double processing gain, one resulting from the periodogram calculation, equal to $10\log_{10}(K \cdot M_{\mathrm{s}})\,$dB \cite{Sturm}, and the other from the beamforming gain, equal to $10\log_{10}(N_\mathrm{R})\,$dB. For this reason, the \ac{MIMO} system can estimate range and speed with high accuracy for \ac{SNR} significantly lower than those reached by the \ac{DoA} estimation algorithm, as it can be seen in Fig.~\ref{figure_a}, Fig.~\ref{figure_b}, and Fig.~\ref{figure_c}. In fact, \ac{MUSIC} is not subject to any processing gain, and the \ac{RMSE} of angle estimation starts to increase at much higher \ac{SNR} values, depending on the number of receiving antennas, $N_\mathrm{R}$. 
In particular, for increasingly negative values of \ac{SNR}, the \ac{RMSE} (in degree) vs \ac{SNR} curves converge approximately to $\Delta \Theta/2.8$, because of the limited search interval, $\Delta \Theta$, over which the pseudo-spectrum is computed. As previously stated, $\Delta \Theta$, and consequently the upper bound value of the curves, strictly depends on the number of antennas, as it can be noticed comparing the blue dashed line and continuous yellow line curves in Fig.~\ref{figure_a}, Fig.~\ref{figure_b} and Fig.~\ref{figure_c}. 

From the estimated range, $\widehat{r}$, and \ac{DoA}, $\widehat{\theta}$, the position estimate of the target is $\widehat{\mathbf{p}}=(\widehat{x}, \widehat{y}) = (\widehat{r} \cos{\widehat{\theta}}, \widehat{r} \sin{\widehat{\theta}})$. The normalized \ac{RMSE}, shown in Fig.~\ref{figure_d}, is derived from the Euclidean distance between the estimated position, $\widehat{\mathbf{p}}$, and the true location of the target, $\mathbf{p}= (x,y) = (r \cos{\theta}, r \sin{\theta})$, divided by the true distance, $r$, as
\begin{equation}
    \text{Normalized \ac{RMSE}} = \sqrt{\frac{1}{N_\mathrm{MC}}\sum_{j=1}^{N_\mathrm{i}}{\frac{\bigl\|{\widehat{\mathbf{p}}}_j - {\mathbf{p}}_j\bigr\|_2^2}{r_j^2}}}.
    \label{eq:norm_RMSE}
\end{equation}
The normalization in \eqref{eq:norm_RMSE} eliminates the dependency of the position \ac{RMSE} on the distance generalizing the results. In fact, since the length of the chord of a circumference is directly proportional to its radius, the \ac{DoA} error causes the position error to increase with distance. In Fig.~\ref{figure_d} the position estimate dependency on the \ac{DoA} is emphasized for different numbers of antennas and 5G numerologies. As the \ac{SNR} decreases, it is possible to notice the impact of \ac{DoA}, which causes a first drop in the performance, and the effect of range estimation error, which leads to a second performance drop at lower \ac{SNR}. 

Another important parameter in sensing is the detection probability $P_\mathrm{D}$ as a function of the \ac{SNR} as shown in Fig.~\ref{figure_e}. In these plots, the threshold $\eta$ has been chosen to ensure a $P_\mathrm{FA} = 1\%$. As expected, since detection is performed on the range-Doppler map, the same map used for velocity and range estimation, it is easy to notice that the range and velocity estimation start degrading when detection probability degrades. Therefore, the main factor limiting the radar performance is \ac{DoA} estimation.
\begin{figure*}[t]
\captionsetup{font=footnotesize,labelfont=footnotesize}
    \centering
    \subfloat[][\label{fig:card_err}]
    {\includegraphics[width=0.4\textwidth]{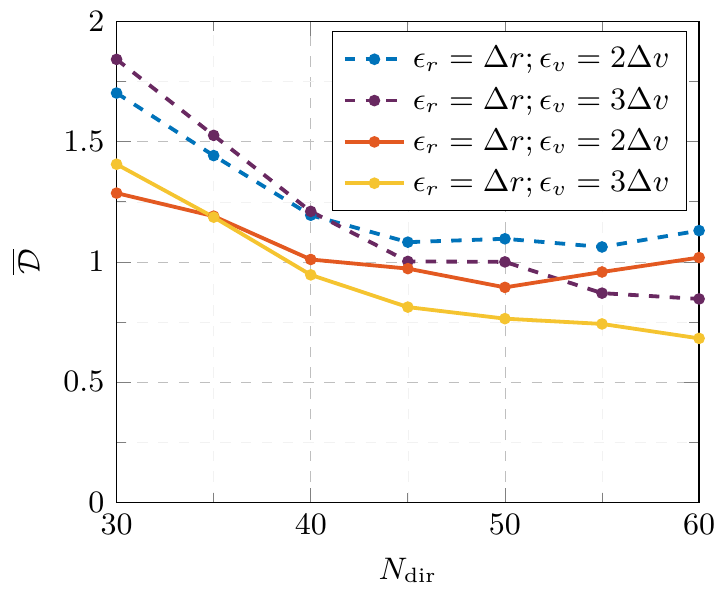}} \qquad
    \subfloat[][\label{fig:loc_err}]
    {\includegraphics[width=0.4\textwidth]{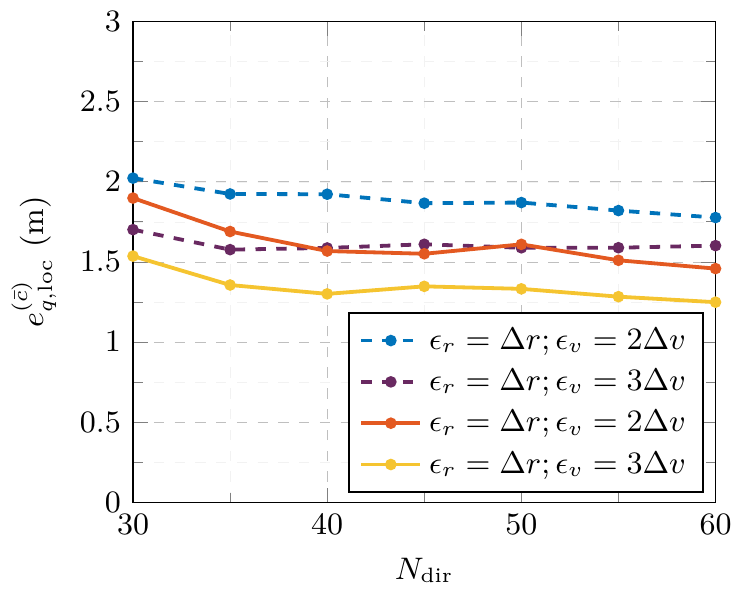}} \\
    \caption{Mean cardinality error (a) and mean \ac{OSPA} localization error (b) varying the number of sensing directions, $N_\mathrm{dir}$, for a \ac{JSC} system with $f_\mathrm{c}=28$GHz, $N_\mathrm{T}=N_\mathrm{R}=50$, obtained with $N_\mathrm{MC}=500$ Monte Carlo iterations. Dashed lines represent the result for $\rho=0.1$, whereas the continuous lines represent the results for $\rho=0.3$.}
    \label{fig:card_loc_err}
\end{figure*}
\subsection{RMSE vs distance} \label{sec:RMSEvsDist}
Let us now analyze the trade-off between communication and sensing varying $\rho$ in \eqref{eq:BFvector}. As it can be observed in Fig.~\ref{figure_3a} and \ref{figure_3b}, the system works well also for moderately low values of $\rho$, e.g., $\rho = 0.1$. Notably, the position \ac{RMSE} is below $0.33\,$m and $0.1\,$m, at $3.5\,\text{GHz}$ and $28\,\text{GHz}$, respectively, even at tens of meters (in \ac{LOS} condition). It is also important to highlight that the \ac{RMSE} values reached at $3.5\,\text{GHz}$ are much higher than that at $28\,\text{GHz}$. This mainly depends on the larger $\Delta \Theta$ resulting from $N_\mathrm{T} = N_\mathrm{R} = 10$, with respect to the beamwidth with $N_\mathrm{T} = N_\mathrm{R} = 50$ antennas. Moreover, at the considered ranges, the \ac{RMSE} of the position mainly depends on \ac{DoA} estimation error, and when this error reaches the upper bound, the position \ac{RMSE} becomes proportional to the distance, $r$, as previously explained in Section~\ref{sec:RMSEvsSNR}.
\begin{figure*}[t]
\captionsetup{font=footnotesize,labelfont=footnotesize}
    \centering
    \subfloat[][\label{fig:OSPA_dist_a}]
    {\includegraphics[width=0.4\textwidth]{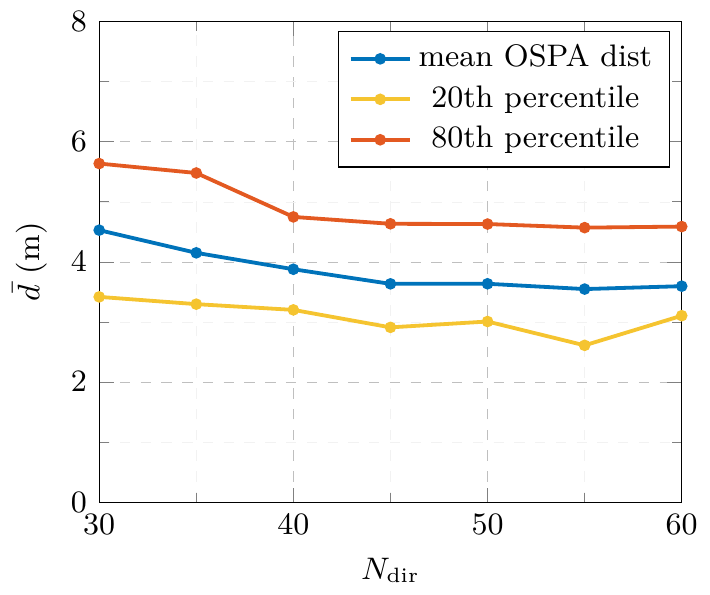}} \qquad
    \subfloat[][\label{fig:OSPA_dist_b}]
    {\includegraphics[width=0.4\textwidth]{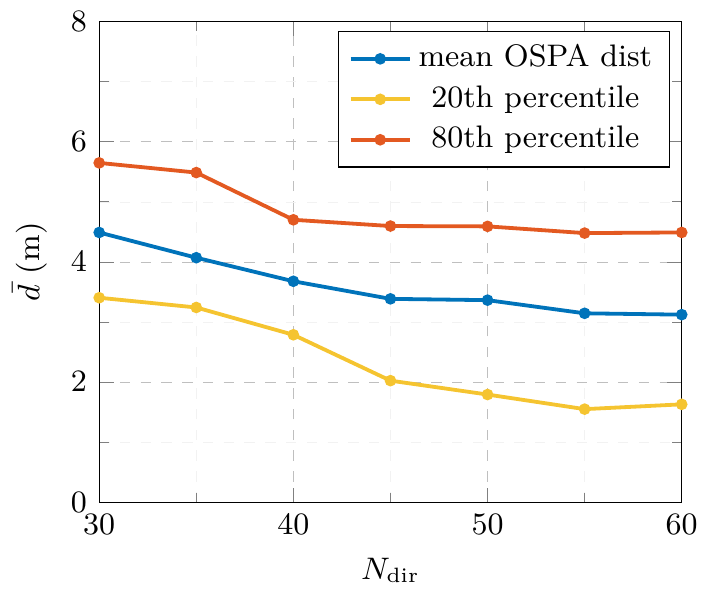}} \\
    \subfloat[][\label{fig:OSPA_dist_c}]
    {\includegraphics[width=0.4\textwidth]{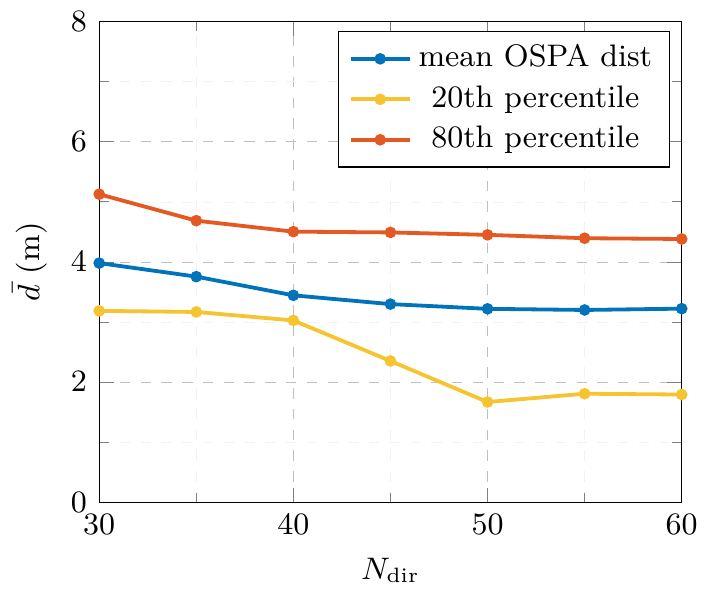}}\qquad
    \subfloat[][\label{fig:OSPA_dist_d}]
    {\includegraphics[width=0.4\textwidth]{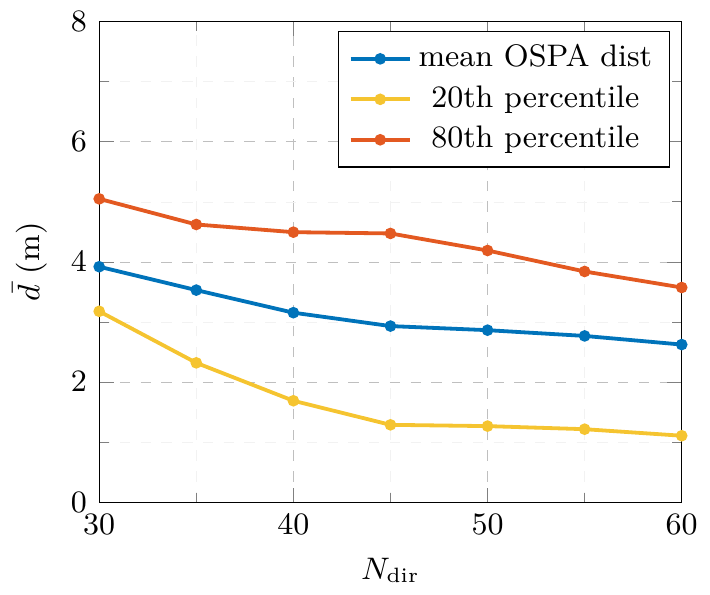}}\\
    \caption{Mean value, $20$th and $80$th percentile of the \ac{OSPA} distance varying the number of sensing directions for different values of $\epsilon_r$, $\epsilon_v$ and $\rho$, with $f_\mathrm{c}=28\,$GHz and $N_\mathrm{T}=N_\mathrm{R}=50$. The \ac{OSPA} metric is computed for $\bar{c}=10\,$m and $q=2$. The considered values are: $\epsilon_r=\Delta r$ and $\epsilon_v=2\Delta v$ with $\rho=0.1$ (a) and $\rho=0.3$ (c), $\epsilon_r= \Delta r$ and $\epsilon_v=3 \Delta v$ with $\rho=0.1$ (b) and $\rho=0.3$ (d).}
    \label{fig:OSPAd_dist}
\end{figure*}
\subsection{Performance analysis of multi-target scenario}
For the multi-object scenario analysis let us consider $L=10$ point targets ($9 + 1$ \ac{UE}). Each target is associated with an \ac{SNR} that depends on its radial distance from the monostatic sensor, on its \ac{RCS}, and on the alignment between the target and the sensing direction, in accordance with \eqref{eq:SNR-path-loss}. Without loss of generality let us assume all the targets with the same \ac{RCS}, equal to $\sigma_\mathrm{RCS}=1\,\text{m}^2$, as in Section~\ref{sec:RMSEvsDist}. The number of \ac{MC} iterations for this group of results is set to $N_{\mathrm{MC}}=500$. In each \ac{MC} iteration targets positions are randomly generated according to a uniform distribution within a sector with radial distance between $20$ to $85\,\text{m}$ and angle from $-60^{\circ}$ to $60^{\circ}$.

The primary purpose of this analysis is to study the performance of the considered \ac{JSC} system when multiple targets are present, computing the \ac{OSPA} metric introduced in Section~\ref{sec:ospa} for different choices of $N_\mathrm{dir}$, ranging from $30$ to $60$ sensing directions. In particular, one of the main objectives is to study the influence of the uncertainty parameters, $\epsilon_r$ and $\epsilon_v$, used in the repeated targets pruning algorithm presented in Section \ref{sec:targPrun}, and of the parameters $\rho$, on the detection and localization capabilities of the system. 

For what concerns $\epsilon_r$ and $\epsilon_v$, a good choice consists of using a multiple of distance and velocity resolutions, $\Delta r$ and $\Delta v$, defined in \eqref{eq:resolution}. In fact, due to the presence of \ac{AWGN}, radial distance and velocity estimates of a repeated target may fall in adjacent bins of the periodogram with respect to those of the original target. In this sense, between range and velocity, it is the latter that presents greater \ac{RMSE} in the low \ac{SNR} regime; this is due to zero padding, which increases velocity resolution at the expense of sensitivity to noise. The mean cardinality error is used to choose these parameters. This metric is given by
\begin{equation}
\mathcal{\overline{D}}=\frac{1}{N_\mathrm{MC}}\sum_{i=1}^{N_\mathrm{MC}}|L-\widehat{L}|.
\end{equation}

\noindent Importantly, this metric does not distinguish between miss-detection, false alarm, and repeated target; however, it can be considered a good indicator for analyzing the average performance of the considered algorithm. In fact, fixing the system parameters, miss-detection and false alarm rates can be regarded as approximately constant, so if $\mathcal{\overline{D}}$ decreases, that should be caused by a drop in the targets' repetition rate.
Fig.~\ref{fig:card_loc_err} shows the mean cardinality error and the mean \ac{OSPA} localization error computed varying the number of sensing directions, $N_\mathrm{dir}$, for different values of $\epsilon_r$, $\epsilon_v$ and $\rho$. As it can clearly be noticed, by fixing $\epsilon_r$ and $\epsilon_v$, the overall performance of the system (both in localization and cardinality error) improves for increasing values of $\rho$. For what concerns the localization error, the results shown in Fig.~\ref{fig:loc_err} are in agreement with those presented in Fig.~\ref{figure_3b}. As the position of the target is varied between $20$ and $85\,\text{m}$, the system performance is worse for $\rho=0.1$ than $\rho=0.3$, as expected.
In Fig.~\ref{fig:card_err} it is possible to notice as for $N_\mathrm{dir} \geq 40$ the mean cardinality error becomes smaller choosing $\epsilon_v = 3 \Delta v$, both for $\rho=0.3$ and $\rho=0.1$, and, in particular, for $\rho=0.3$ the system on average misses less than one target. As pointed out, a value of $\epsilon_r > \Delta r$ does not change appreciably the system performance; therefore, its value is kept fixed, letting $\epsilon_v$ vary. This latter term most affects the repeated targets pruning algorithm detection performance due to zero padding, as already observed. 

In Fig.~\ref{fig:OSPAd_dist}, the mean \ac{OSPA} metric is computed varying $N_\mathrm{dir}$ for the same values of $\epsilon_r$, $\epsilon_v$ and $\rho$ used above. In addition, the $20$th and $80$th percentile are shown to better understand the range of values the \ac{OSPA} metric can assume for different positions of the targets. Also in this case, the best performance are obtained for $\rho=0.3$, $\epsilon_r=\Delta r$ and $\epsilon_v=3\Delta v$. In particular, for this choice of parameters the mean value of $\bar{d}$ is below $3\,\text{m}$ and the $20$th percentile is approximately equal to $1\,\text{m}$ for $N_\mathrm{dir}=60$. Note, however, that such numerical results also consider the portion of the monitored area where the \ac{DoA} estimation is severely degraded (see Fig.~\ref{fig:RMSE_rho_Pd}); a proper sensing cell sizing may avoid such region and lead to much better performance.

\begin{figure*}
\captionsetup{font=footnotesize,labelfont=footnotesize}
    \centering
    \subfloat[][\label{fig:range_angle_map}]
    {\includegraphics[width=0.25\textwidth]{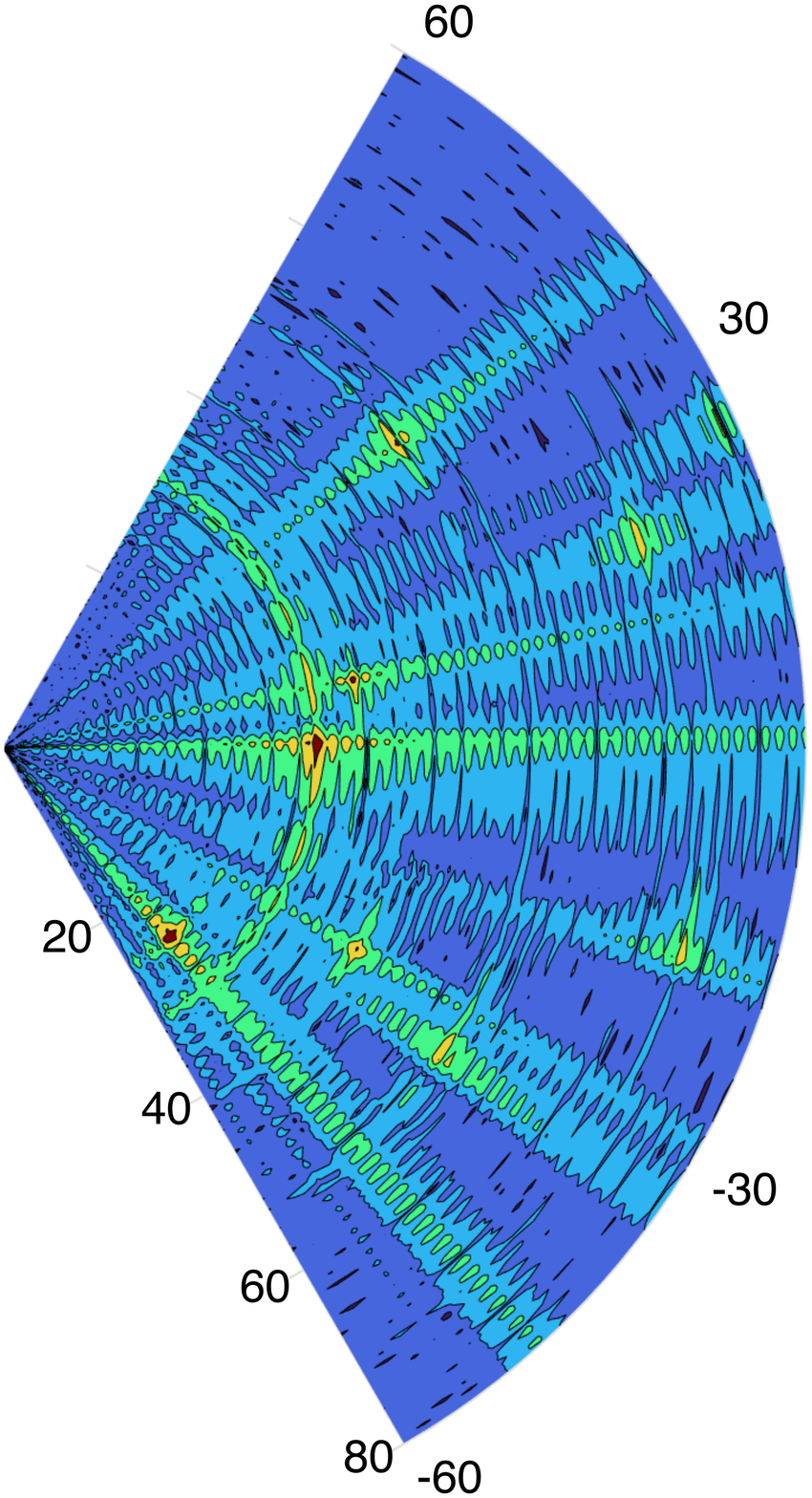}} 
    \subfloat[][\label{fig:range_angle_points1}]
    {\includegraphics[width=0.25\textwidth]{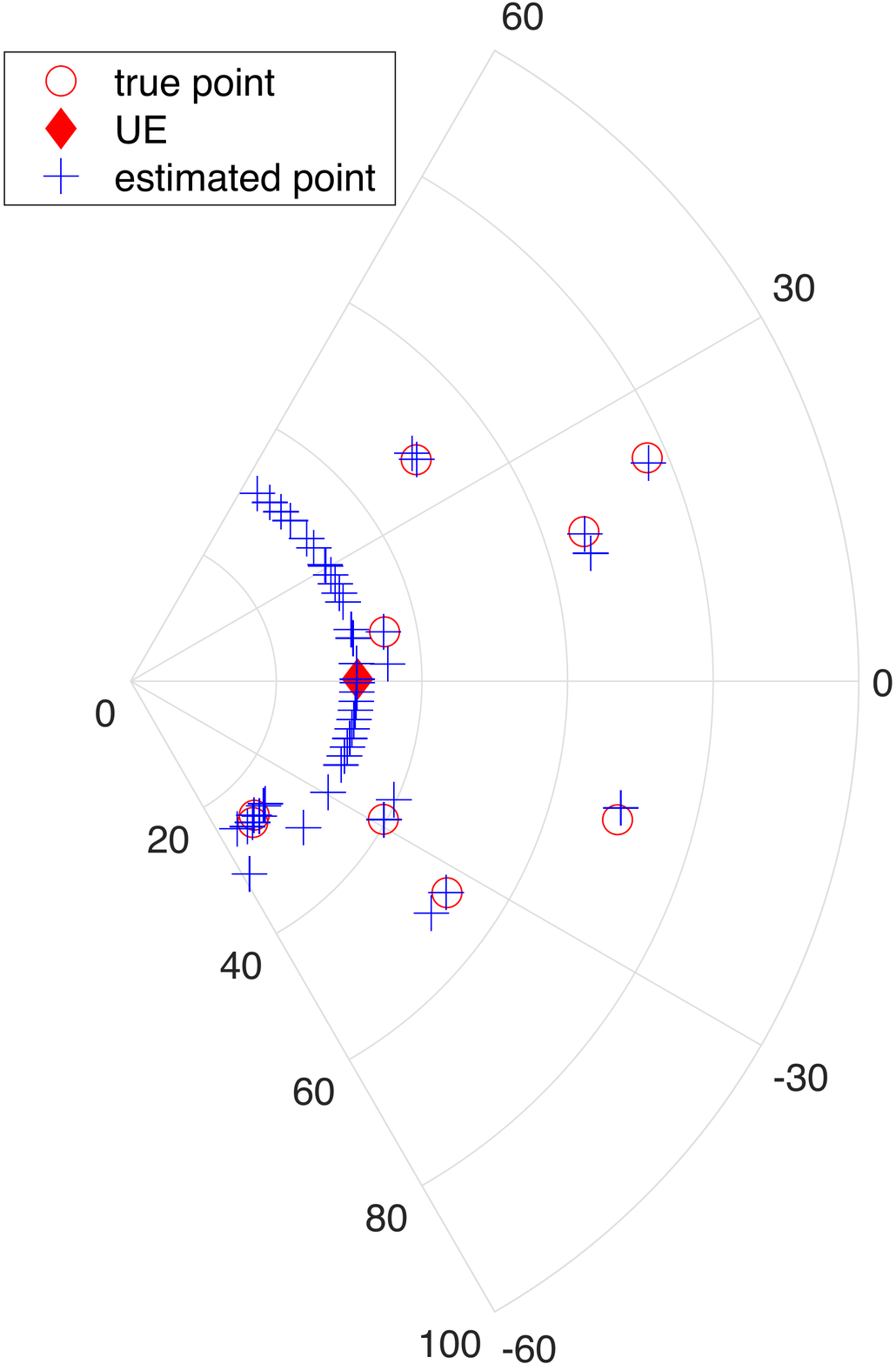}}
      \subfloat[][\label{fig:range_angle_points2}]
    {\includegraphics[width=0.25\textwidth]{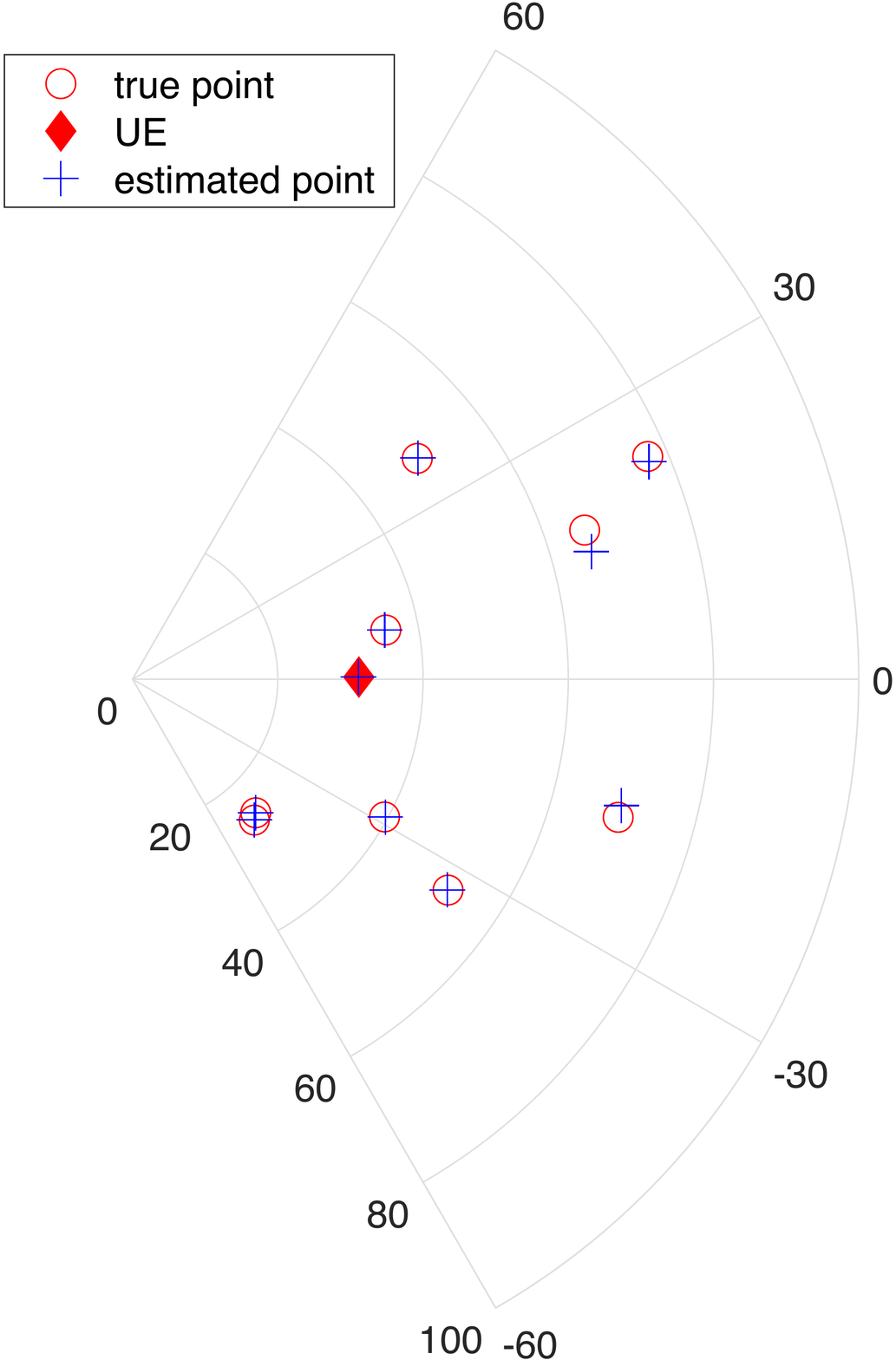}} \\
    \caption{Considered scenario with $9$ targets and $1$ \ac{UE}. The range-angle map in (a) has been obtained with $f_\mathrm{c}=28\,\text{GHz}$, $N_\mathrm{T}=N_\mathrm{R}=50$ antennas, $N_\mathrm{dir}=60$ and $\rho = 0.3$. (b) shows the point detected starting from the range-angle map in (a), before repeated targets pruning. In (c) the result obtained after the removal of the repeated targets, performed with $\epsilon_r = \Delta r$ and $\epsilon_v = 3\Delta v$ is shown.}
    \label{fig:range-angle-map}
\end{figure*}

As a final system-level analysis, in Fig.~\ref{fig:range-angle-map} an example of multiple targets map returned by the \ac{JSC} sensor is shown. The parameters are $\rho=0.3$, $\epsilon_r=\Delta r$ and $\epsilon_v = 3\Delta v$. First, in Fig.~\ref{fig:range_angle_map} the range-angle map obtained by computing the periodogram \eqref{eq:period} in each sensing direction, is shown. Then, in Fig.~\ref{fig:range_angle_points1} we have the targets detected through the hypothesis test \eqref{eq:hypotest}, and the resulting range estimates \eqref{eq:qp_estimate}-\eqref{eq:rv_estimate} and angle estimates \eqref{eq:theta_hat}. As it can be seen, multiple points per target are present. After the repeated targets pruning algorithm introduced in Section~\ref{sec:targPrun} the resulting targets are shown in Fig.~\ref{fig:range_angle_points2}. As we can observe, in this particular case the algorithm is very effective in removing all redundant points while retaining all the useful points. As expected, most of the repeated points are on the circumference with a radius equal to the distance between the \ac{UE} and the monostatic sensor; this is to be attributed to a large fraction of power used for the communication beam which illuminates the \ac{UE} causing a strong received echo.

\section{Conclusion}\label{sec:conclu}
In this work, we designed a multibeam system for \ac{JSC} based on 5G \ac{NR}, capable of detecting and locating multiple targets. We provided a system-level analysis and proposed an algorithm for pruning phantom targets that arise as a consequence of beam-scanning in the presence of beam sidelobes. We identified the main dominant factors affecting performance and compared two system setups operating at sub-$6\,$GHz, and mmWave frequencies. The findings of this paper have demonstrated that: $i)$ \ac{DoA} estimation is the primary source of error when used to evaluate the target position; $ii)$ even with a relatively small fraction of power devoted to sensing, good localization performance at tens of meters can be achieved in \ac{LOS} even at mmWave: $iii)$ in the sub-$6\,$GHz band targets can be detected at higher distances but with lower accuracy mainly because of the reduced number of antenna elements; $iv)$ Tens of targets can be detected and localized with sub-meter level accuracy when the power for sensing is capable of ensuring reliable \ac{DoA} estimation.

\section*{Acknowledgment}
The authors would like to thank Elisabetta~Matricardi for her valuable contribution and Wen~Xu, Ronald~Boehnke, and Tobias~Laas for their helpful suggestions.

\balance

%
\end{document}